\documentclass[letterpaper]{article} 
\usepackage{aaai25}  
\usepackage{times}  
\usepackage{helvet}  
\usepackage{courier}  
\usepackage[hyphens]{url}  
\usepackage{graphicx} 
\usepackage{natbib}  
\usepackage{caption} 
\usepackage{algorithm}
\usepackage{algorithmic}
\usepackage{amsmath}
\usepackage{multirow}
\usepackage{threeparttable}
\usepackage{amsthm}
\usepackage{booktabs}
\usepackage[inline,shortlabels]{enumitem}
\usepackage{enumerate}
\usepackage{makecell}
\usepackage{tabularx}
\usepackage{amsfonts,amssymb} 
\usepackage{pifont}
\usepackage{array}
\usepackage{graphicx}
\usepackage{subfigure}

\usepackage{rotfloat}

\usepackage{caption}
\usepackage{newfloat}
\usepackage{listings}
\DeclareCaptionStyle{ruled}{labelfont=normalfont,labelsep=colon,strut=off} 
\floatstyle{ruled}
\newfloat{listing}{tb}{lst}{}
\floatname{listing}{Listing}
\title{CoPRA: Bridging Cross-domain Pretrained Sequence Models with Complex Structures for Protein-RNA Binding Affinity Prediction}
\author{
    Rong Han\textsuperscript{\rm 1}\equalcontrib,
    Xiaohong Liu\textsuperscript{\rm 2}\equalcontrib,
    Tong Pan\textsuperscript{\rm 3,\rm 4},
    Jing Xu\textsuperscript{\rm 3,\rm 4},
    Xiaoyu Wang\textsuperscript{\rm 3,\rm 4},
    Wuyang Lan\textsuperscript{\rm 5}, \\
    Zhenyu Li\textsuperscript{\rm 1},
    Zixuan Wang\textsuperscript{\rm 1},
    Jiangning Song\textsuperscript{\rm 3,\rm 4}\footnotemark[2],
    Guangyu Wang\textsuperscript{\rm 5}\footnotemark[2],
    Ting Chen\textsuperscript{\rm 1}\footnote{Corresponding authors.}
}
\affiliations{
    \textsuperscript{\rm 1} BNRist, Department of Computer Science and Technology, Tsinghua University\\
    \textsuperscript{\rm 2} UCL Cancer Institute, University College London\\
    \textsuperscript{\rm 3}Monash Data Futures Institute, Monash University \\
    \textsuperscript{\rm 4} Monash Biomedicine Discovery Institute and Department of Biochemistry and Molecular Biology, Monash University\\
    \textsuperscript{\rm 5} State Key Laboratory of Networking and Switching Technology, Beijing University of Posts and Telecommunications\\
    hanr21@mails.tsinghua.edu.cn, xiaohong.liu@ucl.ac.uk, \\
    Jiangning.Song@monash.edu, guangyu.wang24@gmail.com, tingchen@tsinghua.edu.cn
}

\newcommand{\modelname}{CoPRA}
\newcommand{\coformer}{Co-Former}

\usepackage{bibentry}

\begin{document}

\maketitle

\begin{abstract}
Accurately measuring protein-RNA binding affinity is crucial in many biological processes and drug design. Previous computational methods for protein-RNA binding affinity prediction rely on either sequence or structure features, unable to capture the binding mechanisms comprehensively. The recent emerging pre-trained language models trained on massive unsupervised sequences of protein and RNA have shown strong representation ability for various in-domain downstream tasks, including binding site prediction. However, applying different-domain language models collaboratively for complex-level tasks remains unexplored. In this paper, we propose \modelname{} to bridge pre-trained language models from different biological domains via \underline{Co}mplex structure for \underline{P}rotein-\underline{R}NA binding \underline{A}ffinity prediction. We demonstrate for the first time that cross-biological modal language models can collaborate to improve binding affinity prediction. We propose a \coformer{} to combine the cross-modal sequence and structure information and a bi-scope pre-training strategy for improving \coformer{}'s interaction understanding. Meanwhile, we build the largest protein-RNA binding affinity dataset PRA310 for performance evaluation. We also test our model on a public dataset for mutation effect prediction. \modelname{} reaches state-of-the-art performance on all the datasets. We provide extensive analyses and verify that \modelname{} can (1) accurately predict the protein-RNA binding affinity; (2) understand the binding affinity change caused by mutations; and (3) benefit from scaling data and model size.
\end{abstract}

\begin{links}
\link{Code}{https://github.com/hanrthu/CoPRA}
\end{links}

\begin{figure}[t]
\centering
\includegraphics[width=1\columnwidth]{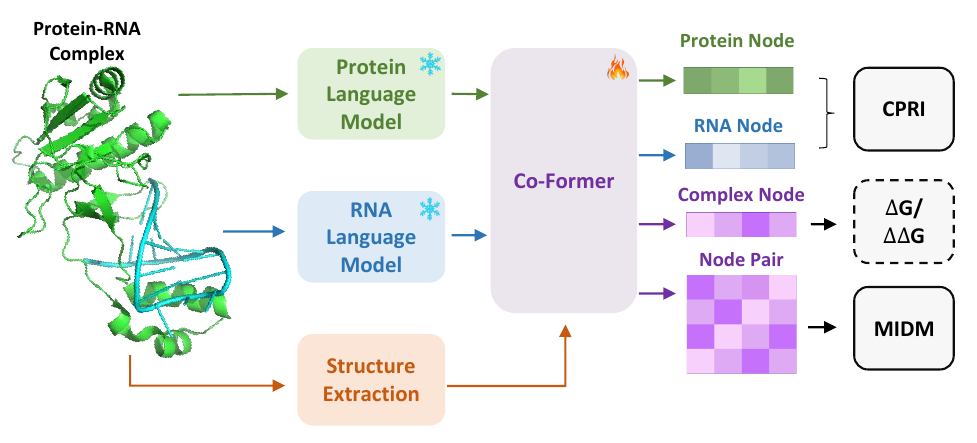} 
\caption{\modelname{} combines Protein and RNA language models with structure information by pre-training on bi-scope tasks with different special embeddings. CPRI: Contrastive Protein-RNA interaction modeling; $\Delta$G/$\Delta\Delta$G: binding affinity/binding affinity change; MIDM: Mask interface distance modeling. The dashed line represents that they are downstream affinity prediction tasks.}
\label{fig1}
\end{figure}

\section{Introduction}

Protein-RNA interactions are crucial in various biological processes, including gene expression and regulation \cite{corley2020rna}, protein translocation, and the cell cycle \cite{zhou2020circular}. Understanding the mechanism of protein-RNA binding is the cornerstone of unraveling complex gene regulatory processes and deciphering the genetic underpinning of diseases, such as neurodegenerative disorders \cite{gebauer2021rna} and kidney disease \cite{seufert2022rna}, leading to the advancement of RNA-based therapies and the design of protein inhibitors that specifically target these interactions.
However, protein-RNA binding is highly flexible. Some proteins bind with RNA through canonical regions while others bind with RNA through intrinsically disordered regions - protein domains characterized by low sequence complexity and highly variable structures \cite{seufert2022rna}, making it challenging to model the mechanism. 

Several computational methods have been proposed for protein-RNA binding affinity prediction, including sequence-based and structure-based methods. The sequence-based approaches process the protein and RNA sequence separately with different sequence encoders \cite{yang2019pnab, pandey_deepnap_2024}, and subsequently model the interactions. However, their performance is often limited because the binding affinity is mainly determined by the binding interface structure \cite{deng_predprba_2019}. Other recent methods are structure-based \cite{hong_updated_2023, harini2024pred}, focusing on extracting structural features at the binding interface, such as energy and contact distance. Based on the extracted features, they developed structure-based machine-learning approaches for affinity prediction. However, these methods are highly dependent on feature engineering with limited generalization ability on new samples due to the limited development dataset size.

Recently, many protein language models (PLMs) \cite{lin2022language, rao2021msa} and RNA language models (RLMs) \cite{penic_rinalmo_2024, chen2022interpretable} have been developed, most of which utilize a mask language modeling strategy \cite{devlin2018bert} to pre-train the models with massive unlabeled sequences. They've shown great performance and generalization ability in various downstream tasks. As the 3D structure of protein/RNA is crucial for understanding their functions, combining structure information into the LMs has recently become a new trend.
For example, SaProt \cite{su_saprothub_2024} pre-trains PLMs with structure information and shows increased performance on different tasks. Instead of adding structure information into pre-training directly, other methods use a much lighter way by combining it with a pre-trained sequence model, such as \cite{jing_crossbind_2024}, showcasing a strong performance gain compared to the sequence-only counterparts. Most of these models are trained and used in single biological modal tasks (i.e. protein or RNA only).

Although the current works show the prosperous potential of structure-informed biological language models for interaction tasks, there are still few works combining pre-trained models from different biological domains. Integrating pre-trained models for multiperspective information extraction has received much attention recently \cite{li2024focusllm}. Modeling cross-modal complex structures for single-modal LMs requires a suitable model design. In the protein-RNA binding affinity prediction task,
one key challenge is the limited size of labeled complex structures, as there are only several datasets that contain a small number of protein-RNA affinity labels, e.g. 135 samples in PRBABv2. Meanwhile, some affinity labels from different datasets may conflict with each other, making it hard to develop and evaluate the models. Therefore, applying different-domain language models collaboratively for complex-level tasks remains less explored.

In this paper, we propose \modelname{}, the first attempt to bridge a PLM and an RLM via \underline{Co}mplex structure for \underline{P}rotein-\underline{R}NA binding \underline{A}ffinity prediction, as shown in Figure \ref{fig1}. Specifically, the overall pipeline of \modelname{} is: 
The protein and RNA sequences are first input into a PLM and an RLM, respectively. Then, we select the embeddings from the two LMs' outputs that are at the interaction interface as the sequence embedding for the subsequent cross-modality learning. The structure information is also extracted from the interaction interface as the pair embedding. We design a lightweight \coformer{} to bridge the interface sequence embedding from two LMs together with the complex structure information. \coformer{} combines the sequence and structure information with a structure-sequence fusion module. We also propose a bi-scope pre-training strategy for \coformer{} to model coarse-grained contrastive interaction classification (CPRI) and fine-grained interface distance prediction (MIDM) at \textbf{atom-wise precision}\footnote{The distance of nodes is by the nearest atom between them.}. 
To deal with the lack of a unified labeled standard dataset issue, we curated the largest protein-RNA binding affinity dataset PRA310 from three public datasets and evaluated \modelname{} and other models' performance. To further demonstrate \modelname{}' ability to understand protein-RNA binding, we adopt it to predict the binding affinity change caused by protein mutation. In summary, our main contributions are listed as follows:
\begin{itemize}[leftmargin=12pt]
    \item We propose \modelname{}, the first attempt to combine protein and RNA language models with complex structure information for protein-RNA binding affinity prediction.
    \item We design a \coformer{} to bridge the embedding of the interface sequence from two LMs together with the complex structure information and design a bi-scope pre-training method, including CPRI and MIDM for understanding the binding from different aspects. \coformer{} is trained on our curated unsupervised dataset PRI30k.
    \item We curate the largest protein-RNA binding affinity dataset PRA310 from multiple data sources. And evaluate the model's performance on three datasets. \modelname{} reaches state-of-the-art performance on multiple datasets, including PRA310 and its subset PRA201 for binding affinity prediction, and a mCSM blind-test set for mutation effect on binding affinity prediction.
\end{itemize}

\begin{figure*}[t]
\centering
\includegraphics[width=0.9\textwidth]{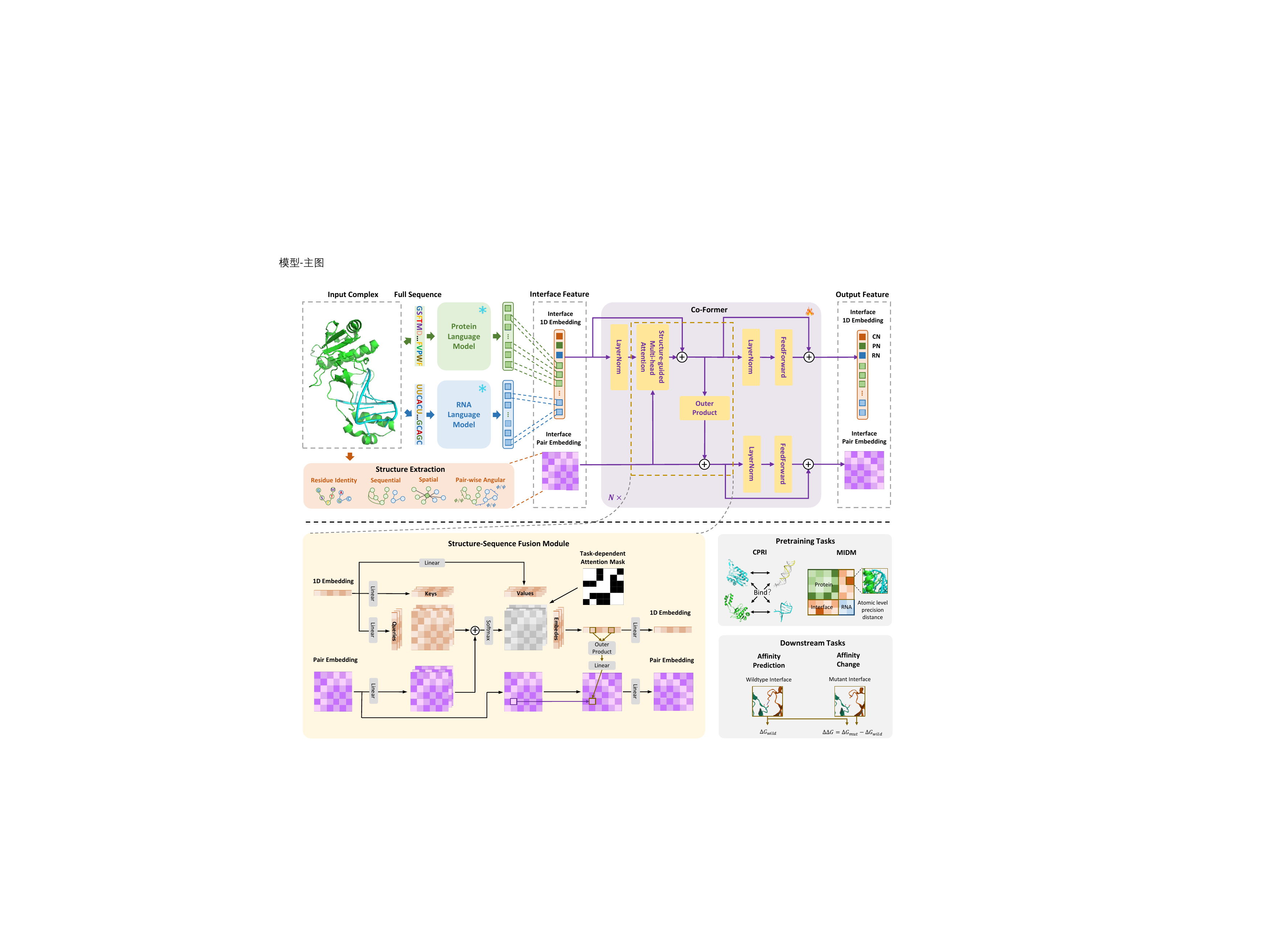} 
\caption{Overview of \modelname{}. Given a protein-RNA complex as input, the sequence information of protein and RNA are fed into a PLM and an RLM, respectively. The output embeddings are selective with interface information and are fed into \coformer{} with pairwise information. The \coformer{} fuses the 1D and pair embedding by structure-guided multi-head attention and outer product modules, with a task-dependent attention mask. The output special nodes and pair embedding of \coformer{} are employed dependent on different tasks, including two pre-training tasks and two downstream affinity tasks. CN, PN, and RN are the special nodes for complex, protein, and RNA, respectively.}
\label{fig2}
\end{figure*}

\section{Related Work}
\subsection{Protein-RNA Binding Affinity Prediction}
Several sequence- or structure-based machine learning-based methods have been applied to predict protein-RNA binding affinity. For example, PNAB \cite{8982930} is a stacking heterogeneous ensemble framework based on multiple machine learning methods, e.g. SVR and Random Forest. They manually extract different biochemical features from the protein and RNA sequences. DeePNAP \cite{pandey_deepnap_2024} is another sequence-based method, leveraging 1D Convolution networks for feature extraction. PredPRBA and PRdeltaGPred \cite{deng_predprba_2019, hong_updated_2023} employ interface structure features for better prediction. Besides, PRA-Pred \cite{HARINI2024129490} is a multiple linear regression model, which utilizes protein-RNA interaction information as features in addition to the protein and RNA information. These studies demonstrate that the sequence feature of RNA/protein, and the interface structure feature both contribute to more accurate prediction. However, most of them only employ part of the information, and it is demanding to develop a method to leverage both sequence and interface structure information.

\subsection{Protein and RNA Language Models} 
Many efforts have emerged to develop foundation language models to leverage the massive biological sequence data. One of the first papers is ESM-1b \cite{rives2021biological} trained on 250 million protein sequences with a BERT-style strategy. Several other PLMs are proposed and perform well on various downstream tasks\cite{rao2021msa,elnaggar2021prottrans,brandes2022proteinbert}. Especially, ESMFold \cite{lin2022language} and OmegaFold \cite{wu2022high} show the power of PLMs on protein structure prediction, without multiple sequence alignment information as in AlphaFold2 \cite{jumper2021highly}. The PLM from ESMFold is named ESM-2, which contains various parameter sizes, from 8M up to 15B. Meanwhile, most RLMs employ a similar paradigm of that in PLMs. The RLMs are trained on massive non-coding RNA sequences. RNA-FM \cite{chen2022interpretable}, Uni-RNA \cite{wang2023uni} and RiNaLMo \cite{penic_rinalmo_2024} are three representative RLMs. They show great ability in RNA function and secondary structure prediction. While PLMs and RLMs have succeeded in many biological tasks, applying them together remains an unexplored area of research.

\subsection{Multi-Modal Learning in Language Models}
Learning from multiple modals can provide the model with multi-source information of the given context \cite{huang2021makes}. Multi-modal learning achieves impressive performance improvement compared to its single-modal counterparts and brings new applications\cite{luo2024vividmed,li2023blip}. Contrastive learning is one efficient unsupervised way to align multi-modal representation to the same semantic space. CLIP \cite{radford2021learning} used an in-batch contrastive strategy to train visual encoders with the text encoders. BLIP-2 \cite{li2023blip} introduces a lightweight QFormer for visual-language pretraining with frozen image encoders and LLMs. In the field of protein, many efforts have been made to integrate the 3D structure information into PLMs. LM-design \cite{zheng2023structure} adds a structure adapter to ESM-2, enabling the structure-informed PLMs on conditional protein design. Recently, SaProt \cite{su_saprothub_2024} and ESM-3 \cite{hayes_simulating_2024} pretrain the PLM with protein sequence and its structural information, increasing the models' overall performance. Existing multi-modal PLMs were trained with the protein structure and sequence modals. It is still an open problem for combining multiple biological modals (e.g. protein and RNA) with complex structure information for complex-level interaction tasks.

\section{Methods}
In this section, we introduce the details of \modelname{}. First, we introduce the overview of \modelname{} and some notations of the protein-RNA complex. Next, we present \coformer{} for bridging the multi-modal information from protein and RNA. Later, we will describe the pre-training task design, including CPRI and MIDM. At last, we will introduce the formulation of downstream tasks, including binding affinity prediction and mutation effect on affinity change prediction. The overall workflow of \modelname{} is described in Figure \ref{fig2}.

\subsection{\modelname{} overview}
\modelname{} is designed to leverage the PLM and RLM for binding affinity prediction. Given a protein-RNA complex $C$, we input the sequence of each protein chain $P_i$ into a PLM, and each RNA chain $R_i$ into an RLM. We generate a sequence embedding and a pair embedding at the binding interface for \coformer{}. The \coformer{} performs structure-sequence fusion and outputs multi-level representations. To develop \coformer{}'s multi-modal understanding, we propose a bi-scope pre-training approach, including CPRI and MIDM, enhancing the model's understanding of protein-RNA complex in different granularity.

\subsection{Notations of the protein-RNA complex}
The input is a protein-RNA complex with at least one protein chain and one RNA chain each. We define the protein as a set of chains $P = \{P_1, ..., P_n\}$, and RNA as $R = \{R_1, ..., R_n\}$. 

\paragraph{Protein.}
Each protein chain contains 1D sequence information $p_i$ and 3D structure information $X_i$ as input, noted as $P_i = \{p_i, X_i\}$. For a chain of length $L_p$, we have $p_i \in \mathbb{A}_p^{L_p}$ and $X_i \in \mathbb{R}^{L_p\times k \times 3}$, where $\mathbb{A}_p$ is the alphabet of protein residue types, including 20 normal amino acids and an unknown token `X'. And $k$ is the number of atoms for representation, we have $k=4$ for \modelname{} modules, containing backbone atoms $\{N, C_A, C, O\}$.

\paragraph{RNA.}
The input of an RNA chain of length $L_r$ is similar to that of proteins, noted as $R_i = \{r_i, X_i\}$, where $r_i \in \mathbb{A}_r^{L_r}$ and $X_i \in \mathbb{R}^{L_r\times k' \times 3}$. The alphabet $\mathbb{A}_r$ of RNA contains only 4 types of base types $\{A, G, C, U\}$ and an unknown token `\_'. Here $k'=4$ for \modelname{} modules, containing backbone atoms $\{P, C_4', C_1', N_1\}$ for pyrimidine base types $\{C, U\}$ and $\{P, C_4', C_1', N_9\}$ for purine base types $\{A, G\}$.

\paragraph{Protein-RNA complex.}
The protein-RNA complex includes sequence and structure information of each chain, and a complex distance map $D$, noted as $C = \{P, R, D\}$. $D \in \mathbb{R}^{L\times L}$, where $L$ is the total node number of the complex. $D$ is generated by full-atom geometry to get the precise pair-wise distance between nodes.

\subsection{Protein-RNA interface representation}
Given a protein-RNA complex input $C=\{P, R, D\}$, we describe here the process for preparing protein-RNA interface representation for \coformer{}. In general, \coformer{} takes a mixed representation at the binding interface, noted as $C_I=\{S, Z\}$. $S \in \mathbb{R}^{(n+3)\times d_s}$ and $Z \in \mathbb{R}^{(n+3)\times (n+3)\times d_z}$, where n is the interface size, $d_s$ is the sequence embedding size and $d_z$ is the pair embedding size.

\paragraph{Interface sequence embedding.}
The full sequence of $P$ and $R$ are fed into PLM and RLM separately to get the full sequence embedding. We select $n$ nodes near the interface according to $D$. Moreover, we design three special nodes as different-level representation aggregators, including a complex node $C^s$, a protein node $P^s$, and an RNA node $R^s$. $C^s$ can attend to all nodes, while $P^s$ can only attend to nodes from proteins and $R^s$ can only attend to nodes from RNAs. These special nodes are randomly initialized and concatenated in front of the interface node embeddings to form $S = C^s\bigoplus P^s\bigoplus R^s\bigoplus P^n\bigoplus R^n$, where $P^n$ and $R^n$ are embeddings for each protein and RNA node.

\paragraph{Interface structure extraction.} Given the interface nodes' (including special nodes) positions of the complex $C$, we can extract invariant pair-wise structure embeddings for \coformer{}. We initialize the special nodes' positions at the geometric center of the interface, the protein, and the RNA, respectively. Inspired by Invariant Point Attention (IPA) in AlphaFold2 \cite{jumper2021highly} for protein feature extraction, we extract four types of pair-wise features from backbone atoms of the complex, including node pair type feature, relative sequential position, distance information, and angular information. More details can be found in Appendix A. The pair-wise information is fed into an embedding layer to form $Z$. As we take the backbone atom positions, $Z$ is unchanged when mutation affects the sidechain conformation. 

\subsection{\coformer{}}
\coformer{} is an N-block dual-path transformer. Each block contains a Structure-Sequence Fusion module SSF, a layer normalization module $\mathrm{LN}$, and a feed-forward module $\mathrm{FFN}$. The form of the $l^{th}$ block is $\{S^{(l+1)}, Z^{(l+1)}\}$ = $\{\mathrm{FFN}(\mathrm{LN}(\hat{S}^{(l)})), \mathrm{FFN}(\mathrm{LN}(\hat{Z}^{(l)}))\}$, where \{$\hat{S}^{(l)}$, $\hat{Z}^{(l)}\}= \mathrm{SSF}(\{S^{(l)}, Z^{(l)}\})$. In this section, we will describe the SSF module in detail.

The SSF module consists of two components, a structure-guided multi-head self-attention module and an outer-product update module, as shown in Figure \ref{fig2}. Given the $l^{th}$ layer's input $\{S^{(l)}, Z^{(l)}\}$, the pair embedding $Z^{(l)}$ is first projected to the head size and added to the attention embedding, guiding attention with structural information. Then, we take a pair-wise outer product for the updated sequence embedding ${\hat{S}^{(l)}}$ to get the pair embedding $\hat{Z}^{(l)}$. The module can be formulated as:

\begin{align}
    &Q^{(l)},K^{(l)},V^{(l)} = S^{(l)}[W^{(l)}_Q, W^{(l)}_K, W^{(l)}_V], \\
    &A^{(l)} = \frac{Q^{(l)}K^{(l)T}}{\sqrt{d_k}} + \mathrm{Linear}(Z^{(l)}), \\
    &\hat{S}^{(l)} = (\mathrm{Softmax}(A^{(l)}) \odot M) \cdot V^{(l)},\\
   &o^{(l)}_{ij} = \hat{s}^{(l)}_i \otimes \hat{s}_{j}^{(l)T}, \\
   &\hat{z}_{ij}^{(l)} = z^{(l)}_{ij} + \mathrm{Linear(o^{(l)}_{ij})},
 \label{eq1}
\end{align}

where, $W_Q, W_K$ and $W_V$ are the projection matrices, and $M$ is the task-dependent mask, as detailed in Figure \ref{fig4} in Appendix B. $\hat{s}^{(l)}_i, \hat{s}^{(l)}_j \in \mathbb{R}^{1\times d_s}$ are the $i^{th}$ and $j^{th}$ feature of $\hat{S}^{(l)}$. $\hat{s}^{(l)}_i \otimes \hat{s}^{(l)T}_j$ is the outer product, resulting in $o^{(l)}_j \in \mathbb{R}^{d_s\times d_s}$. $z^{(l)}_{ij}$, $\hat{z}^{(l)}_{ij}$ is from position $(i, j)$ of $Z^{(l)}$ and $\hat{Z}^{(l)}$, respectively. We simplify the multi-head attention in the equation for easy understanding.

\subsection{Bi-scope Pre-training}
In this section, we will describe the pre-training tasks, including a cross-modal contrastive protein-RNA interaction (CPRI) task for understanding interaction pairs (whether they interact) and a mask interface distance modeling (MIDM) for understanding the atom-precision node distance (how they interact) given only backbone structure as input.

\paragraph{Contrastive interaction modeling.}
Utilizing protein and RNA representations for cross-modal matching is similar to image-text matching. We formulate this problem in an in-batch way, inspired by CLIP \cite{radford2021learning}. Specifically, for a protein $P$ and an RNA $R$ from a complex $C$,  we mask the interface structure information of the pair embedding $Z$ and get the output protein and RNA special node embedding from \coformer{}, denoted as $P^s$, $R^s$. Given a batch of protein-RNA complexes of batch size $K$, we generate $K^2$
pairs $(P^s_i, R^s_j)$, where $i, j \in \{1,...,K\}$. The pair is positive when $i=j$ and the other pairs are negative. We adopted a symmetric contrastive loss function for the training:
\begin{align}
   \mathcal{L}_i^P(P_i^s, \{R_j^s\}_{j=1}^{K})&=-\frac{1}{K}\mathrm{log}\frac{\mathrm{exp}(s(P_i^s, R_i^s)/\tau)}{\sum_{j}\mathrm{exp}(s(P_i^s, R_j^s)/\tau)},\\
  \mathcal{L}_i^R(R_i^s, \{P_j^s\}_{j=1}^{K})&=-\frac{1}{K}\mathrm{log}\frac{\mathrm{exp}(s(R_i^s, P_i^s)/\tau)}{\sum_{j}\mathrm{exp}(s(R_i^s, P_j^s)/\tau)},\\
  \mathcal{L_{CPRI}} &= \frac{1}{2}\sum_{i=1}^{K} (\mathcal{L}_i^P + \mathcal{L}_i^R),
 \label{eq2}
\end{align}

where, $s$ denotes the similarity of the embeddings and we adopt cosine similarity in practice, and $\tau$ is the temperature.

\paragraph{Mask interface modeling.}
Modeling the atom-precision distance is crucial for understanding how the protein-RNA nodes interact. We design a coarse- to fine-grained pre-training method. Specifically, as described in Figure \ref{fig4} in Appendix B, 50\% of the pair embedding $Z$ will be masked with a ratio of 15\%, and the other 50\% will be unchanged. The model is required to reconstruct the interface distance. The ground truth distance of two nodes is defined by the nearest atoms from each node, thus the model needs to infer the interface detail from sequence embedding and partially masked pair embedding $Z$. \textbf{All} the distance at the \textbf{interface} will be used for loss calculation. To make the training more stable, we divide the distance into multiple bins, where the bins at the close part are dense and at the remote part are sparse, converting the task into a classification task with a cross-entropy loss:
\begin{align}
   O_i &= \mathrm{Interface}(\mathrm{Linear}(Z_i^{(N)})),\\
   \mathcal{L}_{MIDM, i} &=-\frac{1}{L^2} \sum_{j, k=1}^L\mathrm{log}\frac{\mathrm{exp}(o_{ijk,t}/\tau)}{\sum_{b}\mathrm{exp}(o_{ijk,b}/\tau)}y_{ijk,t}, 
 \label{eq3}
\end{align}

where, $O_i$ is the distance prediction of the $i^{th}$ complex, and $y_{ijk, t}$ is the distance for the $i^{th}$ complex at position $(j, k)$, with the label $t$, and $\tau$ is the temperature. With a hyperparameter $\alpha$, the in-batch bi-scope pre-training loss is :
\begin{align}
       \mathcal{L} = \mathcal{L}_{CPRI} + \alpha \cdot (\frac{1}{K}\sum_{i=1}^{K} \mathcal{L}_{MIDM, i}). 
       \label{eq4}
\end{align}

\subsection{Protein-RNA affinity prediction tasks}
The downstream tasks consist of protein-RNA binding affinity prediction and protein mutation effect on binding affinity prediction. Here is the formulation of these two tasks. We take $\mathrm{MSE}$ loss for both tasks.
\paragraph{Binding affinity prediction.} Given a complex $C$ as input, we fed the output special node's embedding $C^s$ into an MLP to predict $\Delta G$, noted as $\Delta G = \mathrm{MLP}(C^s)$.

\paragraph{Mutation effect on binding affinity prediction
.} This task predicts the binding affinity change between the mutant and the wild complex\footnote{This is the common representation, while in mCSM, the label is defined as $\Delta\Delta G=\Delta G_{wild} - \Delta G_{mut}$.}, noted as $\Delta\Delta G = \Delta G_{mut} - \Delta G_{wild}$. Since \coformer{} only requires backbone structure information, we can input the same backbone structure and different sequences to get $C_{wild}^s, C_{mut}^s$, making it convenient for prediction, note as $\Delta\Delta G = \mathrm{MLP}(C^s_{mut}) - \mathrm{MLP}(C^s_{wild})$.

\section{Experiments}

\begin{table*}[ht]
\setlength{\belowcaptionskip}{0.3cm}

\centering
\small
\begin{tabular}{l|ccc|cccc|cccc}
\Xhline{1pt}
\multirow{2}{*}[-0.5ex]{Method} & \multirow{2}{*}[-0.5ex]{Struc} & \multirow{2}{*}[-0.5ex]{Seq} & \multirow{2}{*}[-0.5ex]{LM} &\multicolumn{4}{c|}{PRA310}  & \multicolumn{4}{c}{PRA201}                                           \\ 

\multicolumn{1}{c|}{} & \multicolumn{3}{c|}{}  & \multicolumn{1}{c}{RMSE$\downarrow$} & \multicolumn{1}{c}{MAE$\downarrow$} & \multicolumn{1}{c}{PCC$\uparrow$} & \multicolumn{1}{c|}{SCC$\uparrow$} & \multicolumn{1}{c}{RMSE$\downarrow$}                    & \multicolumn{1}{c}{MAE$\downarrow$} & \multicolumn{1}{c}{PCC$\uparrow$} & SCC$\uparrow$           \\ \midrule
LM+LR & \ding{55} & \ding{51} & \ding{51} & 1.801 & 1.472 & 0.365 & 0.348 & 1.750 & 1.383 & 0.370 & 0.362 \\

LM+RF &\ding{55} & \ding{51} & \ding{51} & 1.561 & 1.248 & 0.418 & 0.457 & 1.569 & 1.252 & 0.437 & 0.467 \\
LM+MLP &\ding{55} & \ding{51} & \ding{51}& 1.688 & 1.388 & 0.412 & 0.428 & 1.638 & 1.282 & 0.403 & 0.412 \\
LM+SVR &\ding{55} & \ding{51} & \ding{51}&  1.506 & 1.209 & 0.475 & 0.489 & 1.476 & 1.192 & 0.454 & 0.456 \\ 
LM+Transformer &\ding{55} & \ding{51} & \ding{51} & 1.481 & 1.192  & 0.489 & 0.485 & 1.433 & 1.172 & 0.492 & 0.487 \\
DeepNAP* \cite{pandey_deepnap_2024} &\ding{55} & \ding{51} & \ding{55}& - & - & - & - & 1.964 & 1.600 & 0.345 & 0.349 \\ 
\cmidrule{1-12}

PredPRBA* \cite{deng_predprba_2019} &\ding{51} & \ding{55} & \ding{55} & - & - & - & - & 2.238 & 1.695 & 0.370 & 0.316 \\
FoldX$^\dag$ \cite{delgado2019foldx} &\ding{51} & \ding{55} & \ding{55} &   - &  -  & 0.212 & 0.283 & - & - &  0.212  &  0.268  \\
GCN \cite{kipf2016semi}  &\ding{51} & \ding{55} & \ding{55}  & 1.705  & 1.378 & 0.145 & 0.144  & 1.631 & 1.322 & 0.201   & 0.203     \\
GAT \cite{velivckovic2017graph}  &\ding{51} & \ding{55} & \ding{55}  & 1.644  & 1.337 & 0.238 & 0.174  & 1.542 & 1.235 & 0.262 & 0.221    \\
EGNN \cite{satorras2021n} &\ding{51} & \ding{55} & \ding{55}  & 1.634  & 1.340   &  0.226 & 0.212 & 1.639 & 1.345 &    0.241   &    0.217   \\
GVP \cite{jing2020learning} &\ding{51} & \ding{55} & \ding{55}  & 1.678  & 1.361   &  0.262 & 0.283 & 1.702 & 1.372 &    0.240   &    0.305   \\
IPA \cite{jumper2021highly} &\ding{51} & \ding{55} & \ding{55}  & 1.462 & 1.208 & 0.495 & 0.496 & 1.464 & 1.191 & 0.510 & 0.514\\

LM+IPA &\ding{51} & \ding{55} & \ding{51} & 1.454 & 1.198  & 0.514 & 0.496 & \underline{1.405} & \underline{1.159} & 0.532 & 0.507 \\ \cmidrule{1-12}

\modelname{} (scratch) &\ding{51} & \ding{51} & \ding{51}& \underline{1.446} & \underline{1.188} & \underline{0.522} & \underline{0.520} & 1.428 & 1.172 & \underline{0.534} & \underline{0.526} \\
\modelname{} &\ding{51} & \ding{51} & \ding{51}& \textbf{1.391} & \textbf{1.129} & \textbf{0.580} & \textbf{0.589} & \textbf{1.339} & \textbf{1.059} & \textbf{0.569} & \textbf{0.587} \\

\Xhline{1pt}

\end{tabular}

\caption{The mean metrics of 5-fold cross-validation on the PRA310 and PRA201 datasets. Sequence- and structure-based models are listed in the table. * They only provide a web server with input requirements, so we only test them on the PRA201 subset. $^\dag$ The FoldX prediction is the complex energy change whose value is much larger, thus we only compare the correlation coefficient. LM is ESM-2 + RiNALMo. CoPRA (scratch) represents CoPRA without pre-training on CPRI and MIDM. The standard deviation can be found in the Appendix E.} 
\label{Table main}
\end{table*}

\subsection{Exeriment Setup}
\paragraph{Pre-training dataset.} The pre-training dataset used here is curated by ourselves, capturing protein-RNA pairs of multiple poses. There are in total 5,909 protein-RNA complexes in the Protein Data Bank (PDB), which were collected in a pair-wise form in BioLiP2. They define each interacting protein-RNA chain pair in the complex as an entry, resulting in 150k chain pairs. We create the non-redundant pre-training dataset PRI30k with the annotation of BioLiP2 by finding the maximum connected subgraph in each complex. More details can be found in Appendix C.
\paragraph{Affinity datasets.} Existing affinity datasets only contain a small number of protein-RNA affinity data with inconsistent labels across datasets. It is necessary to build a standard dataset for benchmarking. We collect samples from three public datasets, PDBbind \cite{wang2004pdbbind}, PRBABv2 \cite{hong_updated_2023}, and ProNAB \cite{harini_pronab_2022}. After removing duplication we get 435 unique complexes. We carefully compare the inconsistent labels from the raw literature and calibrate the annotations. We then filter complexes with length and chain number limits, resulting in 310 complexes. We name our dataset PRA310, which is the largest and most reliable dataset under the same settings. We utilize CD-HIT \cite{fu2012cd} to generate complex clusters, with a sequence identity threshold of 70\%. We split these clusters for a standard 5-fold cross-validation setting. PRA201 is a subset of PRA310, containing only one protein chain and one RNA chain in each complex with a stricter length limit. More details can be found in Appendix C. The mCSM blind test set is a dataset from mCSM \cite{pires2014mcsm}, containing 79 non-redundant single-point mutations from 14 protein-RNA complexes.

\paragraph{Metrics and implementation details.} Following \cite{pandey_deepnap_2024}, we take 4 metrics for evaluation, including the root mean square error (RMSE), the mean absolute error (MAE), the Pearson correlation coefficient (PCC), and the Spearman correlation coefficient (SCC). We take ESM-2 650M \cite{lin2023evolutionary} and RiNALMo 650M \cite{penic_rinalmo_2024} as our LMs. All the experiments are conducted on 4 NVIDIA A100-80G GPUs. The block number of \coformer{} is 6, with a sequence and pair embedding size of 320 and 40, respectively. In pre-training, we set MIDM's mask ratio to 15\%. We use the Adam optimizer with an initial learning rate of 3e-5. The node number of the interface is 256. The introduction of baselines is in Appendix D.

\subsection{Predicting Protein-RNA Binding Affinity}
We first evaluate our model's performance on PRA310 and PRA201. We divide the baseline methods into sequence- and structure-based. As illustrated in Table \ref{Table main}, the scratch version of \modelname{} reaches the best performance on the PRA310 dataset. IPA is the best-performed model without LMs, and we replace the sequential input of IPA with the embeddings from LMs, improving its performance with 0.19 on PCC. Moreover, most methods with LM embedding as input perform better than others, indicating the great power of combining pre-trained unimodal LMs for affinity prediction. We then pre-train our model with PRI30k, increasing the overall performance significantly on both datasets. On PRA310, \modelname{} gets an RMSE of 1.391, MAE of 1.129, PCC of 0.580, and SCC of 0.589, much better than the second-best model \modelname{} (scratch). The PredPRBA and DeepNAP only provide web servers and support protein-RNA pair affinity prediction, and we compared the methods on the PRA201 dataset with them. Although at least 100 samples in PRA201 appear in their training set, their performance on PRA 201 is significantly lower than that they reported, indicating the less generalization ability of these methods. This phenomenon can be explained by the experiment of PRdeltaGPred \cite{hong_updated_2023} that removes worst-performed samples, as shown in Appendix E. Moreover, we observe a consistent performance improvement of most models from PRA310 to PRA201, indicating that PRA310 is more comprehensive and challenging. The experiments in PRA310 and PRA201 show \modelname{}'s ability to precisely predict the binding affinity, especially when equipped with the proposed bi-scope pre-training.
\begin{table}[ht]
    \centering
    \small

    \begin{tabular}{lcccc}
        \toprule
        Method & RMSE$\downarrow$ & MAE$\downarrow$ & PCC$\uparrow$ & SCC$\uparrow$  \\
        \midrule
        FoldX (zero-shot) &1.727&1.496&\textbf{0.474}&\textbf{0.548}\\
        \modelname{} (zero-shot)  & \textbf{0.994} & \textbf{0.737} & 0.314 & 0.411 \\
        \midrule
        DeepNAP* & 1.106 & 1.004 & 0.428 & 0.339  \\
        mCSM  & 1.814 & 1.478 & 0.528 & 0.466  \\
        \modelname{}   & \textbf{0.957} & \textbf{0.833} & \textbf{0.550} & \textbf{0.570} \\
        \bottomrule
    \end{tabular}
    \caption{Per-structure performance on mCSM blind test set. * DeePNAP's training set overlaps with this test set.}
    \label{table 2}
\end{table}

\subsection{Predicting Mutation Effects on Binding Affinity}
To further evaluate our model's understanding of affinity in a fine-grained way, we redirect our model to predict the protein's single-point mutation effect on the protein-RNA complex. Following works in protein mutation effects prediction \cite{luo_rotamer_2023}, the metrics are averaged at a \textbf{per-complex} level. We evaluate both zero-shot and fine-tuned performance of \modelname{}, after pre-training on PRI30k and tuning on PRA310. As shown in Table \ref{table 2}, Notably, ours (zero-shot) has a competitive performance, outperforming other models under the RMSE and MAE metrics. After fine-tuning with the cross-validation set used by mCSM, our model outperforms other models in all four metrics, with RMSE of 0.957, MAE of 0.833, PCC of 0.550, and SCC of 0.570. This superior performance comes from the bi-scope pre-training targets, although not see any mutational complex structures. The performance demonstrates \modelname{}'s generalization ability on different affinity-related tasks.

\subsection{Ablation study}
In this section, we present extensive ablation studies of our model to explore its performance on PRA310, including the module parts, the pertaining strategy, and the model size.
\begin{table}[ht]
    \centering
    \small
    \begin{tabular}{lcccc}
        \toprule
        Method & RMSE$\downarrow$ & MAE$\downarrow$&  PCC$\uparrow$ & SCC$\uparrow$  \\
        \midrule
        \modelname{}   & \textbf{1.391} & \textbf{1.129} & \textbf{0.580} & \textbf{0.589}  \\
        - Pre-train    & 1.446 & 1.188  & 0.522 & 0.520  \\
        - Pair info & 1.454 & 1.177 & 0.518 & 0.519  \\
        - Crop patch  & 1.481 & 1.192  & 0.489 & 0.485  \\
        - Special nodes  & 1.497 & 1.211  & 0.456 & 0.469  \\
        - \coformer{} & 1.688 & 1.388 & 0.412 & 0.479 \\
        \bottomrule
    \end{tabular}
    \caption{Ablation study on modules}
    \label{table 3}
\end{table}

\paragraph{Modules ablation.} We \textbf{progressively delete} the modules of \modelname{}. As shown in Table \ref{table 3}, removing each component of \modelname{} will cause a performance decrease, demonstrating the necessity and importance of the modules we designed. The removal of pre-training causes a significant loss of performance, indicating that our pre-training strategy is crucial for affinity prediction. However, the removal of pair information from the scratch version of \modelname{} does not cause a significant loss of performance while removing the patch cropping will cause an obvious decrease. Because the interface information can help the model directly, adding more information on top of the interface cropping may not be helpful when the sample number is limited and the binding mode is flexible. The special nodes also increase the model's performance because they are indeed different levels of attention-based readout functions, effective for multi-level representation of the complex. If we remove all the components and only feed the LMs' output into an MLP, the performance will be much poorer, thus brutally combining embeddings without a suitable model is impracticable.

\begin{table}[ht]
    \centering
    \small

    \begin{tabular}{lcccc}
        \toprule
        Method & RMSE$\downarrow$ & MAE$\downarrow$ & PCC$\uparrow$ & SCC$\uparrow$ \\
        \midrule
        Scratch              & 1.446 & 1.188 & 0.522 & 0.520  \\
        CPRI             & 1.442 & 1.165  & 0.528 & 0.522  \\
        DM            & 1.445 & 1.167 & 0.542 & 0.535  \\
        CPRI+DM      & 1.418 & 1.167 & 0.558 & 0.541  \\
        CPRI+IDM & 1.421 & 1.142 & 0.560 & 0.542  \\
        CPRI+MIDM         & \textbf{1.391} & \textbf{1.129}     & \textbf{0.580} & \textbf{0.589} \\
        \bottomrule
    \end{tabular}
    \caption{Ablation study on pretraining strategy}
    \label{table 4}
\end{table}

\paragraph{Pretraining strategy ablation.} Based on Table \ref{table 4}, we can observe that when only trained with one pre-training target, the distance modeling (DM) brings better performance than CPRI. This is because the distance modeling task is more fine-grained and provides more information for affinity tasks. Combining CPRI with various DM tasks improves the overall performance. Moreover, the results suggest that distance at the interface is more important than that within protein and RNA, thus directly modeling the interface is a better strategy. After masking some of the pair embeddings, the task becomes more challenging, urging the model to get an in-depth understanding of the relationship between the node type and distance.

\begin{figure}[htbp]
    \subfigure[Correlation]{
       \centering
        \includegraphics[width=0.23\textwidth]{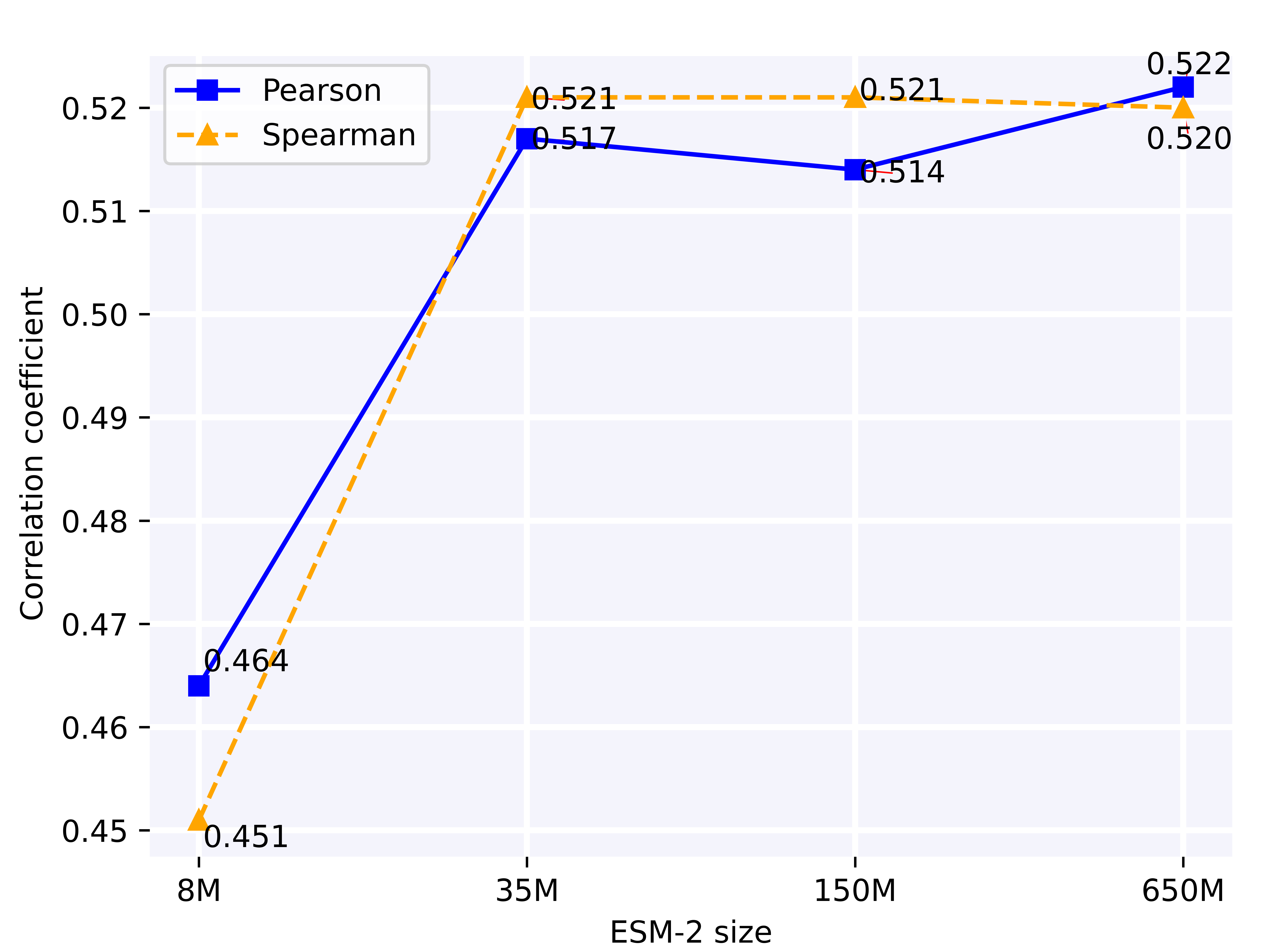}
    }\subfigure[Error]{
        \centering
        \includegraphics[width=0.23\textwidth]{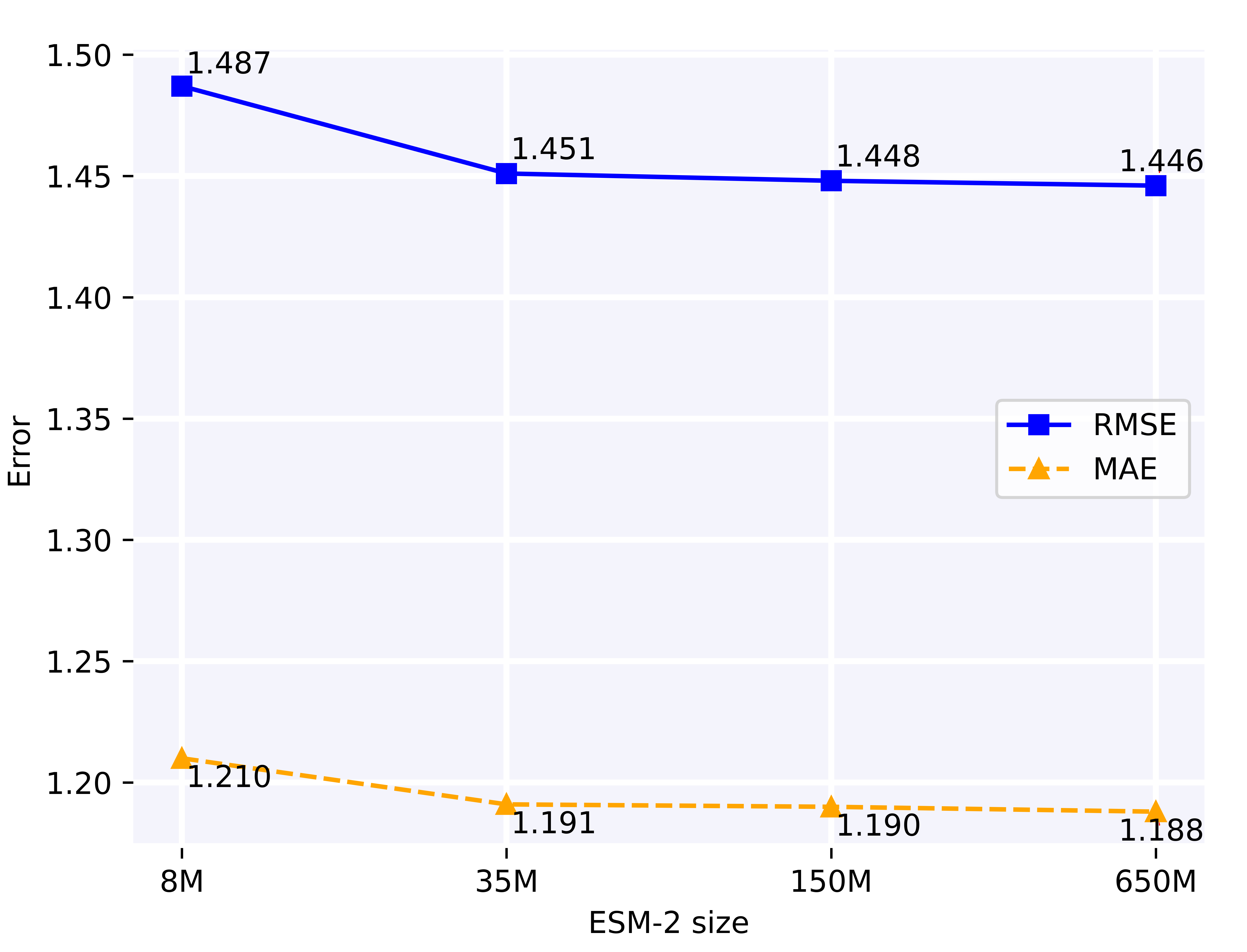}
    }\caption{Ablation study of ESM-2 model size.}
    \label{fig3}
\end{figure}

\paragraph{Model size ablation}

Since RiNALMo only provides a 650M pre-trained model, we ablate the size of ESM-2 and train our \modelname{} from scratch. As shown in Figure \ref{fig3}, increasing the model size brings improvement in performance, and the best-performed model is the ESM-2 650M model. This is because larger pre-trained models can provide larger embedding dims, containing better representation ability gained from unsupervised sequences. This is consistent with the performance trends observed in the unimodal language models. We demonstrate that when doing cross-modal tasks, the collaborator model's size is also of important consideration, and the larger model will probably result in better complex-level performance.

\section{Conclusion}
In this work, we present \modelname{}, the first attempt to combine different biological language models with structural information for protein-RNA binding affinity prediction. We design a \coformer{} for sequence and structure feature fusion and propose an effective bi-scope pre-training approach. Meanwhile, we curate the largest standard protein-RNA binding affinity dataset PRA310 for 5-fold cross-validation and a pre-training dataset PRI30k. Our model achieves state-of-the-art performance on the two binding affinity datasets and the mutation effect prediction dataset. 

Future research can extend the model to more biological domains, such as protein-DNA binding affinity prediction. While our model performs well in predicting the protein's single-point mutation effect on the complex, it is also important to extend the application to multi-point mutation and RNA mutations.

\section{Acknowledgments}
This work is supported by the National Key R\&D Program of China (2021YFF1201300, 2021YFF1201303, 2022YFC2703105), National Natural Science Foundation of China (grants 62272055), New Cornerstone Science Foundation through the XPLORER PRIZE, Guoqiang Institute of Tsinghua University, and Beijing National Research Center for Information Science and Technology (BNRist). We also acknowledge financial support from the Major and Seed Inter-Disciplinary Research projects awarded by Monash University (J.S.). The funders had no roles in study design, data collection and analysis, the decision to publish, or manuscript preparation.

\bibliography{aaai25}

\newpage
\appendix

\section{Supplementary Materials}

\section{A. Structure Extraction Approaches}
Inspired by IPA \cite{jumper2021highly}, we implement a similar invariant pair embedding $Z$ extraction approach from a protein-RNA complex. Specifically, the invariant structure information comes from four aspects and the feature is calculated pairwisely, including residue identity, sequential relative distance from each chain, spatial distance, and angular information. The feature at the $(i, j)$ position of $Z$ is denoted as $z_{ij}$. The details are described as follows:

\paragraph{Node identity pair embedding.} There are 29 node types, including 20 normal amino acids, 1 unknown amino acid, 4 base units, 1 unknown base unit, and 3 special nodes. The special nodes include a protein node, an RNA node, and a complex node, as described in the Methods section. For each node pair $i$ and $j$, the residue-pair identity is $t_{ij} \in \mathbb{N}^{29^2}$. The residue-pair embedding is denoted as $E_r = \mathrm{Embed}_r(T)$. The identity pair embedding is asymmetric because it is order-aware.

\paragraph{Relative sequential embedding.} The relative sequential information is defined within each chain. If two nodes are from different chains, this information will be omitted. The special nodes are in the same `super chain' and the distance is 1 between each other. The sequential embedding is denoted as $E_s = \mathrm{Embed}_s(D_{seq})$. The relative sequential embedding is symmetric.

\paragraph{Distance embedding.} Given input positions $X \in \mathbb{R}^{L\times k\times 3}$, the distance information between each backbone atom pair of each node pair is calculated as a distance map $D_{pair}\in \mathbb{R}^{L\times L\times k \times k}$. then reshaped to $ D_{pair} \in \mathbb{R} ^ {L \times L\times (k*k)}$. Then, the distance is transformed by a Gaussian kernel and we input this pair distance to get $E_d = \mathrm{Embed}_d(D_{pair})$. The distance embedding is symmetric.

\paragraph{Angular embedding.} Given input positions $X \in \mathbb{R}^{L\times k\times 3}$, two dihedral angles $\phi, \psi$ are calculated between each node pair, given the backbone information. As we use 4 atoms as the backbone atoms for protein amino acids and RNA bases, the information is calculated pairwisely and results in $A = \{\phi, \psi\} \in [0, 2\pi)^2$, which is skew-symmetric. The angular embedding is denoted as $E_a = \mathrm{Embed_a}(A)$.

After getting these four embeddings, the pair embedding $Z$ is calculated by $Z = \mathrm{MLP}(\mathrm{Cat}(E_r, E_s, R_d, E_a))$, where $\mathrm{Cat}$ is concatenlation at the embedding dim. Note that the pair embedding $Z$ is invariant, which means rotation and translation of the whole complex will not change $Z$. Moreover, only the embedding at the interface changes if we rotate/translate protein or RNA only.

\section{B. Different task masks}
\begin{figure}[htbp]
    \subfigure[Mask for CPRI]{
       \centering
        \includegraphics[width=0.45\textwidth]{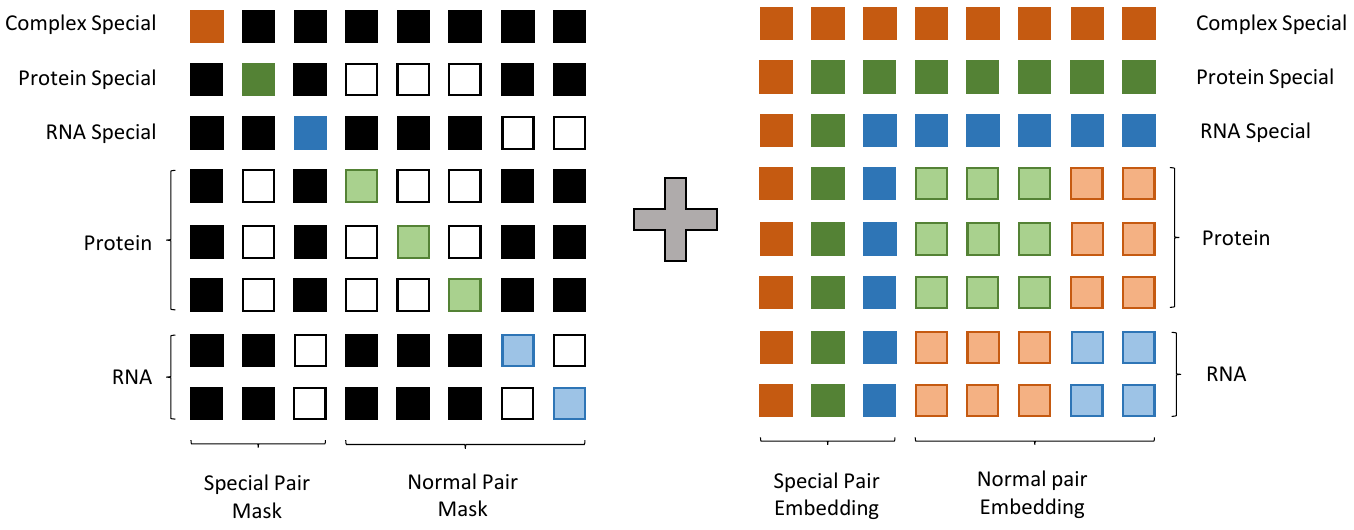}
    } \\
    \subfigure[Mask for MIDM]{
        \centering
        \includegraphics[width=0.45\textwidth]{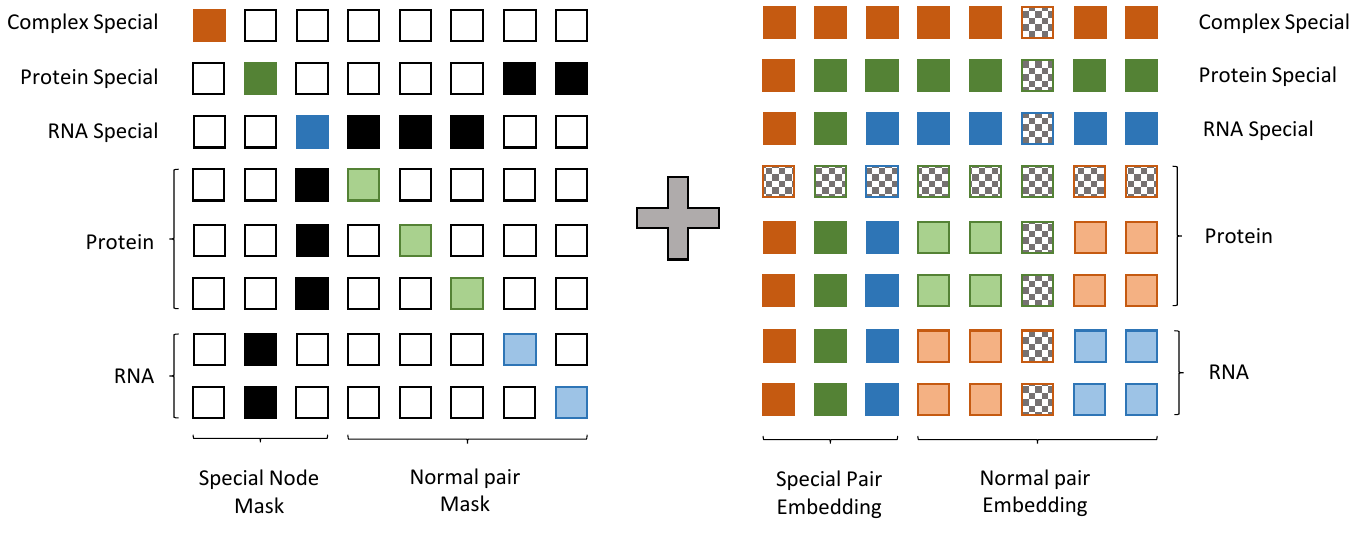}
    } \\
    \subfigure[Mask for Affinity Prediction Tasks]{
        \centering
        \includegraphics[width=0.45\textwidth]{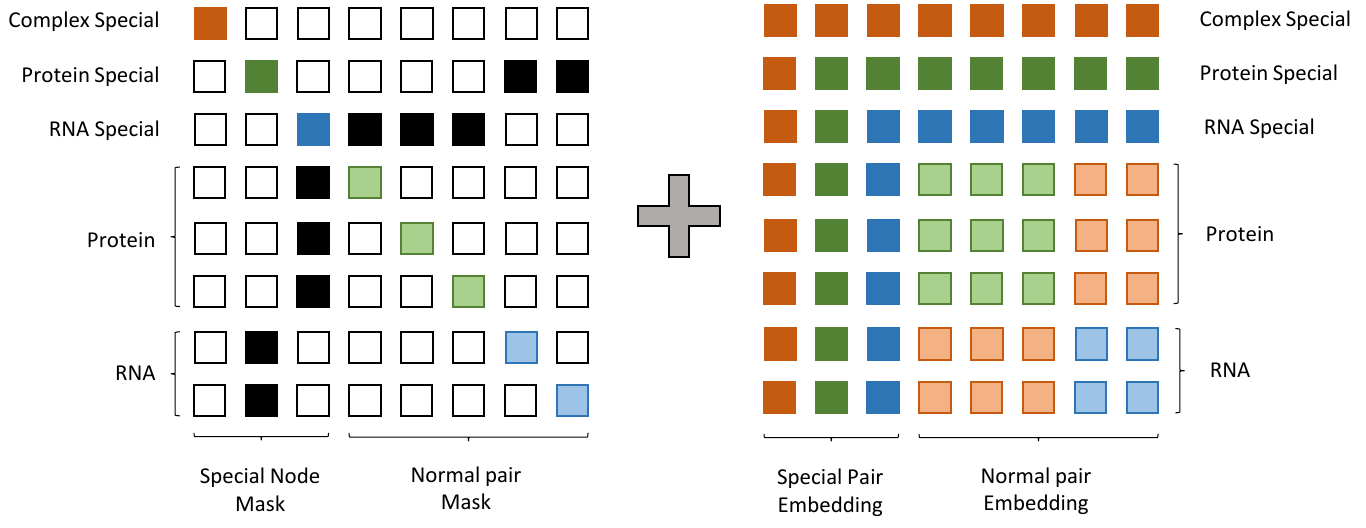}
    } \\
    \caption{Different task masks.}
    \label{fig4}
\end{figure}

In this work, we design two attention mask types and one pair embedding mask for the two pre-training tasks and the downstream tasks, according to the task targets. The mask design is described in Figure \ref{fig4}. For each task, we show the attention mask setting and the input pair embedding.

\paragraph{Mask for CPRI.} The target of CPRI is to classify whether a protein and an RNA interact based on their own sequence and structure information. Therefore, we mask all the positions that may share the structure information of the protein and the RNA. All the nodes can only attend to the nodes from the same type of macromolecules. The pair embedding is calculated normally without masking because they will not pass messages across protein and RNA with the CPRI attention mask. Finally, we only use the special node embeddings from protein and RNA for the contrastive binary interaction classification.

\paragraph{Mask for MIDM.} The attention mask for MIDM has fewer constraints than that for CPRI. Since we hope the model understands the atom-precision distance given the partially masked backbone information, the nodes can attend to all nodes to collect comprehensive information. Therefore, only the protein and RNA special nodes cannot attend to the nodes from different macromolecules. Meanwhile, we mask some nodes' position information, thus all the related pair embedding is replaced with the same mask embedding, as shown in Figure \ref{fig4}(b). We use the pair embedding of all the normal nodes as the output for the distance modeling prediction.

\paragraph{Mask for $\Delta G$ and $\Delta\Delta G$.}
When applied to downstream tasks, we set the attention mask the same as that of MIDM. Each node, except the protein and the RNA special nodes, can attend to each other for interaction. Meanwhile, the pair embedding is also not masked, providing full backbone structure information for the prediction of $\Delta G$ and $\Delta\Delta G$. Finally, the special complex node's embedding is input into an affinity head for affinity predictions.

\section{C. Dataset construction and statistics}
In this section, we will introduce the construction of PRI30k and PRA310 in detail, including the data source, the data construction, the filtering standard, the split strategy, and the final statistics. We provide the community with a high-quality standard dataset with five-fold splits for the evaluation of affinity prediction methods, and a non-redundant pre-training dataset for understanding protein-RNA interactions.

\subsection{PRA310}
\paragraph{Dataset source and filtering.} The data of PRA310 comes from three public datasets, including PDBbind \cite{wang2004pdbbind}, PRBABv2 \cite{hong_updated_2023} and ProNAB \cite{harini_pronab_2022}. We present these three datasets' detailed information before and after filtering in Table \ref{table 5}. After combining the data sources, we get 435 unique samples, and we filter each dataset with a maximum total protein residue length $L_p<= 1000$ a maximum total RNA base length $5 <=L_r <= 500$, and a maximum chain number of 4, resulting in 310 samples. Meanwhile, if we retain the samples with only one protein chain and one RNA chain, there are 201 samples left.

\begin{table}[ht]
    \centering
    \small

    \begin{tabular}{lccc}
        \toprule
        Name & Samples all & Samples filtered & Pair samples \\
        \midrule
        PDBbind & 321 & 210 & 138\\
        PRBABv2  & 145 & 105 & 59\\

        ProNAB  & 338 & 219 & 140\\
        \midrule
        Ours   & 435 & 310 & 201\\
        \bottomrule
    \end{tabular}
    \caption{Statistics of different datasets.}
    \label{table 5}
\end{table}

\paragraph{Merge labels and Relabel conflict annotations.} In total, 39 samples have conflicts of annotations. We manually check the paper sources and correct them. There is also another sample in PDBbind that is reported to be mislabeled in PRBABv2 and we correct its annotation. Therefore, we curate the largest dataset with the most samples and reliable labels, which is suitable for model evaluation and benchmarking.

\paragraph{Five-fold split.} To avoid data leakage, we follow the common practice of splitting clusters according to protein sequence identity. We use CD-HIT \cite{fu2012cd} to cluster the protein sequences with a sequence identity threshold of 70\% to get several chain-level clusters. Since our dataset contains samples of more than one protein chain, we merge the clusters that contain chains from the same clusters to produce complex-level clusters. Finally, the complex clusters are randomly split into five folds with a seed of 2024.

\subsection{PRI30k}
\paragraph{Dataset source and filtering.}
We exhaustively collected all the protein-RNA complexes from the Protein Data Bank, resulting in 5909 samples. We also split the complexes into protein-RNA pairs for our bi-scope pre-training. Since there may be more than one unique protein-RNA binding pair in each complex, we designed a filtering strategy to get the non-redundant interacting pairs. BioLIP2 \cite{zhang2024biolip2} is an up-to-date protein interaction dataset, identifying 150k raw protein-RNA interacting pairs curated from the complexes in the Protein Data Bank. With the annotations from BioLIP2, we can locate the interacting pairs. However, since many protein-RNA complexes are symmetric assemblies, the dataset is highly redundant, and there are many super-long chains, which is not suitable for developing our computing methods. Therefore, We need to create a non-redundant dataset for efficient pre-training. First, we filter the dataset with a maximum protein residue length $L_p<= 750$ and a maximum RNA base length $5 <=L_r <= 500$. Then we designed a rule to find the max connected subgraph from BioLIP2, as described in Algorithm \ref{alg1}. Since most redundant structures are symmetric, we only need to start from the first chain and find a max-connected subgraph containing this chain.

\begin{algorithm}[htb]
\caption{Max connected subgraph (MCS)}
\label{alg1}
\textbf{Input}: $\mathbb{A} = [(C_{p1}, C_{r1}), ... (C_{pn}, C_{rn})]$, $I_{p} = \{C_{p1}: [C_{r1}, ... C_{rk}], ..., C_{pn}: [C_{rx}, ...C_{rn}]\}$,  $I_{r} = \{C_{r1}: [C_{p1}, ... C_{pk'}], ..., C_{rn}: [C_{px'}, ...C_{pn}]\}$\\
\textbf{Output}: Non-redundant pairs $\mathbb{A}'$.
\begin{algorithmic}[1]
\STATE Let $\mathbb{A}' = \{(C_{p1}, C_{r1})\}$.
\STATE Let $\mathbb{A}_{prot} = \{C_{p1}\}$.
\STATE Let $\mathbb{A}_{rna} = \{C_{r1}\}$

\STATE Let search = 1.
\WHILE{search == 1}
\STATE Let search = 0. \\
$\mathbb{A}', \mathbb{A}'_{rna}, \mathrm{search} = \mathrm{Search}(\mathbb{A}', \mathbb{A}_{prot}, \mathbb{A}_{rna}, I_p)$

$\mathbb{A}', \mathbb{A}'_{prot}, \mathrm{search} = \mathrm{Search}(\mathbb{A}', \mathbb{A}_{rna}, \mathbb{A}_{prot}, I_r)$

\ENDWHILE \\
\textbf{return} $\mathbb{A}'$
\end{algorithmic}

\end{algorithm}

\begin{algorithm}[htb]
\caption{Search}
\label{alg2}
\textbf{Input}: $\mathbb{A}', \mathbb{A}_x, \mathbb{A}_y, I_x$\\
\textbf{Output}: $\mathbb{A}', \mathbb{A}'_y$ 
\begin{algorithmic}[1]
\STATE Let search = 0.
\FOR{$C_{xi}$ in $\mathbb{A}_{x}$}
    \FOR{$C_{yi}$ in $I_x[C_{xi}]$}
    \IF{$C_{yi}$ not in $\mathbb{A}_{y}$}
        \STATE Let $\mathbb{A}_{y} = \mathbb{A}_{y} \cup \{C_{yi}\}$ \\
        \STATE Let $\mathbb{A}' = \mathbb{A}' \cup \{(C_{xi}, C_{yi})\}$
        \STATE Let search = 1.
    \ENDIF
    \ENDFOR
\ENDFOR

\textbf{return} $\mathbb{A}', \mathbb{A}'_y, \mathrm{search}$
\end{algorithmic}

\end{algorithm}

\begin{figure}[htbp]
    \subfigure[5WFK]{
       \centering
        \includegraphics[width=0.23\textwidth]{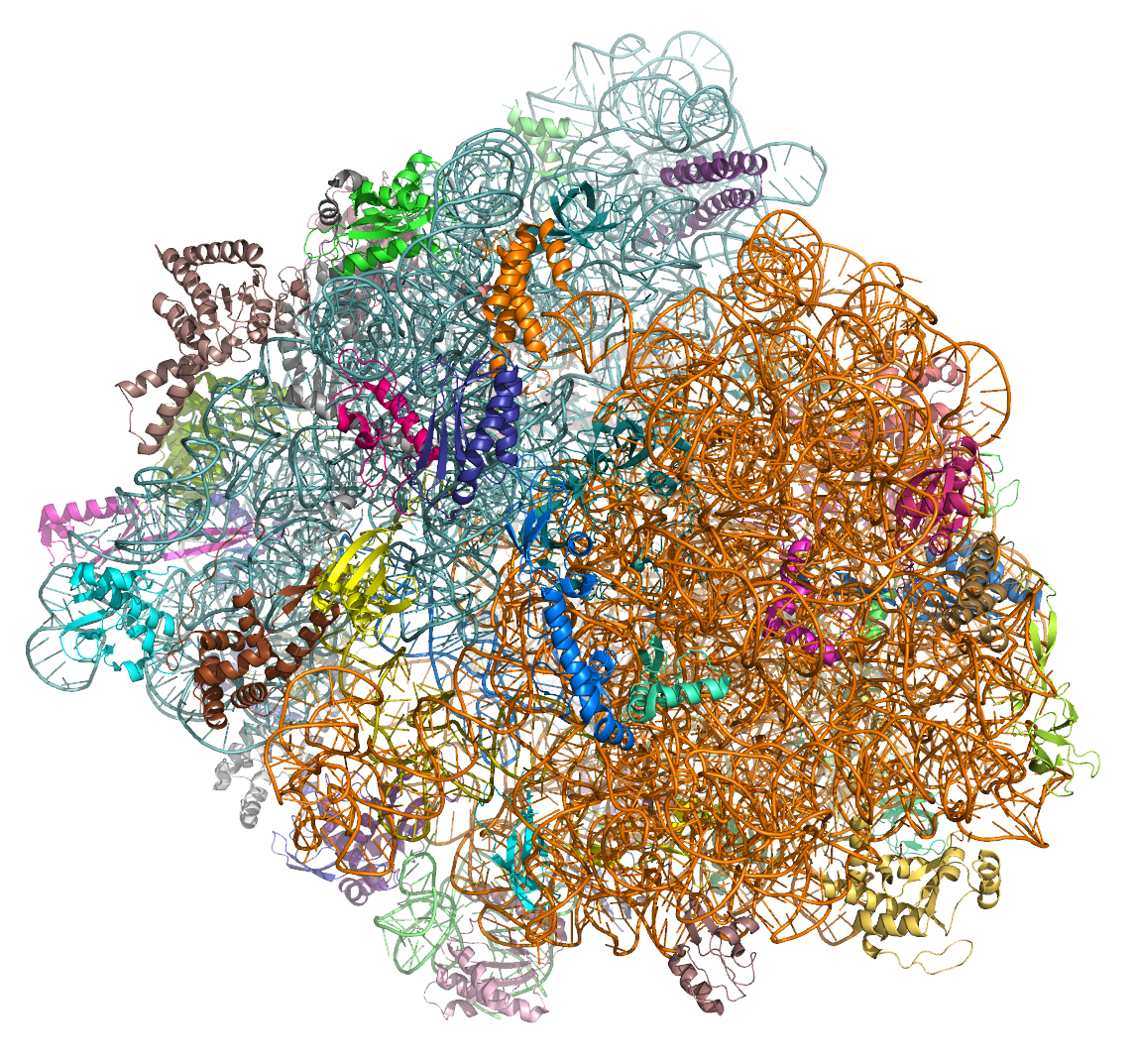}
    }\subfigure[5WFK filtered]{
        \centering
        \includegraphics[width=0.23\textwidth]{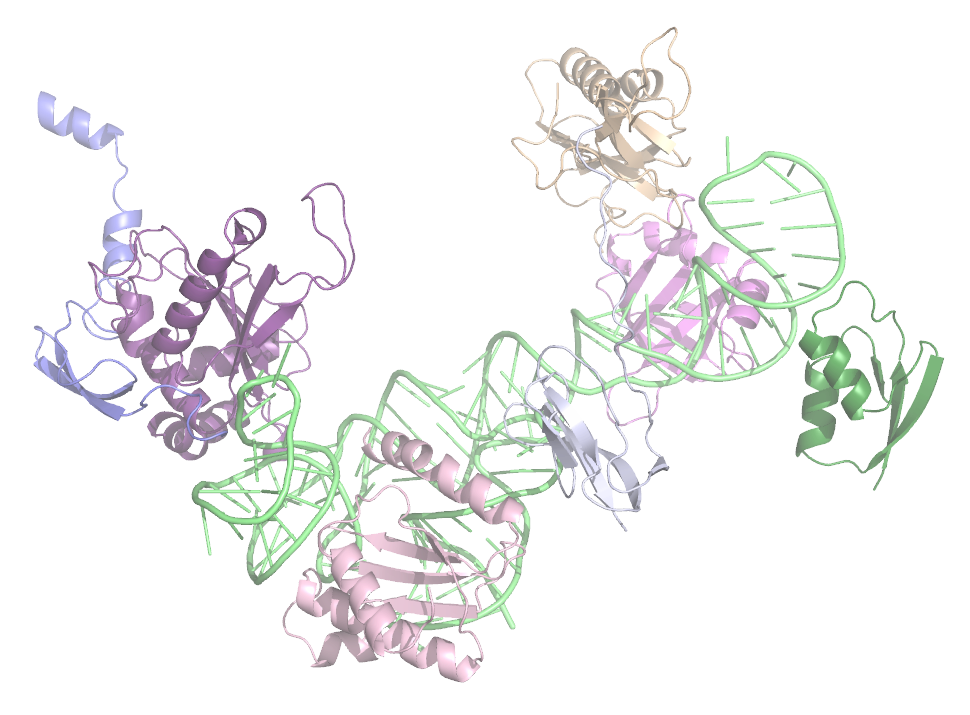}
    }\caption{An example of complex before and after filtering.\label{inut_output}}
    \label{fig5}
\end{figure}

 In the pseudocode, we define each chain as a node, and if two nodes interact, there is an edge between them. After filtering, we can get the non-redundant pairs from each complex, as shown in Figure \ref{fig5}, after filtering a complex (PDB id: 5WFK), we reduce the total chains from 59 chains to 8 interacting chains.

\begin{figure}[htbp]
    \subfigure[Pose 1]{
       \centering
        \includegraphics[width=0.23\textwidth]{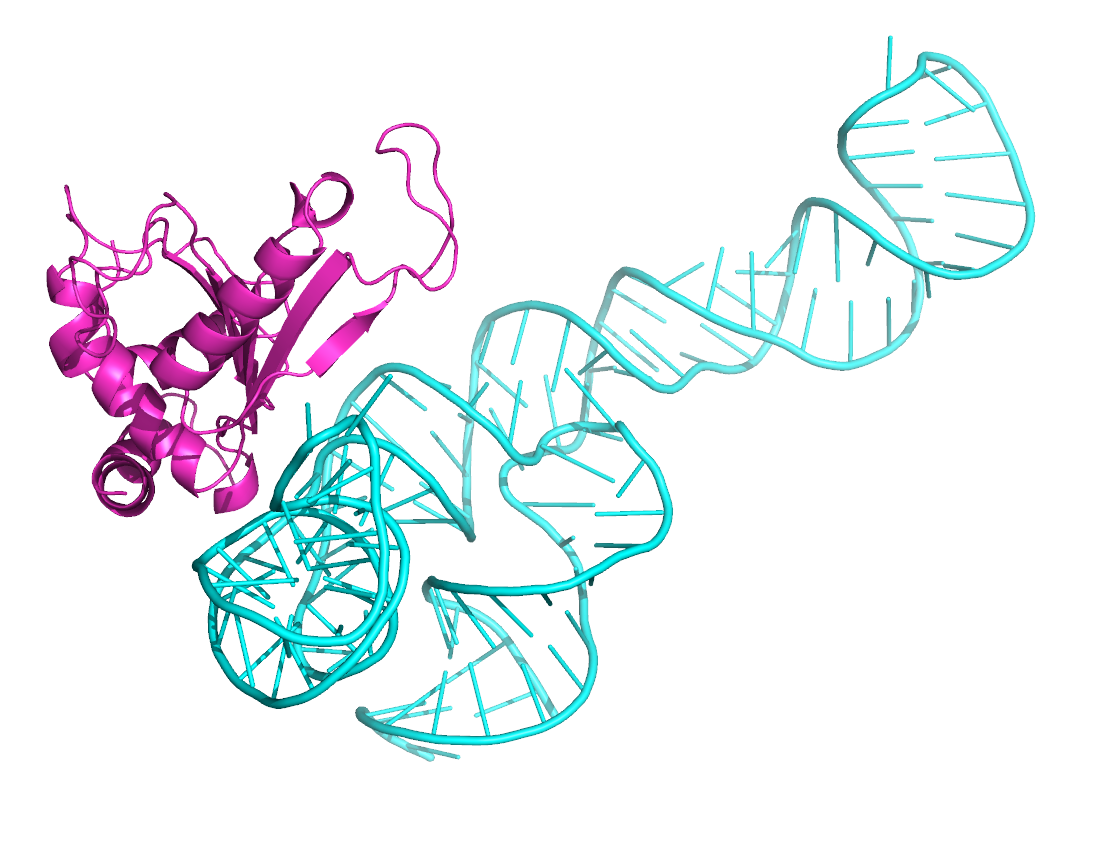}
    }\subfigure[Pose 2]{
        \centering
        \includegraphics[width=0.23\textwidth]{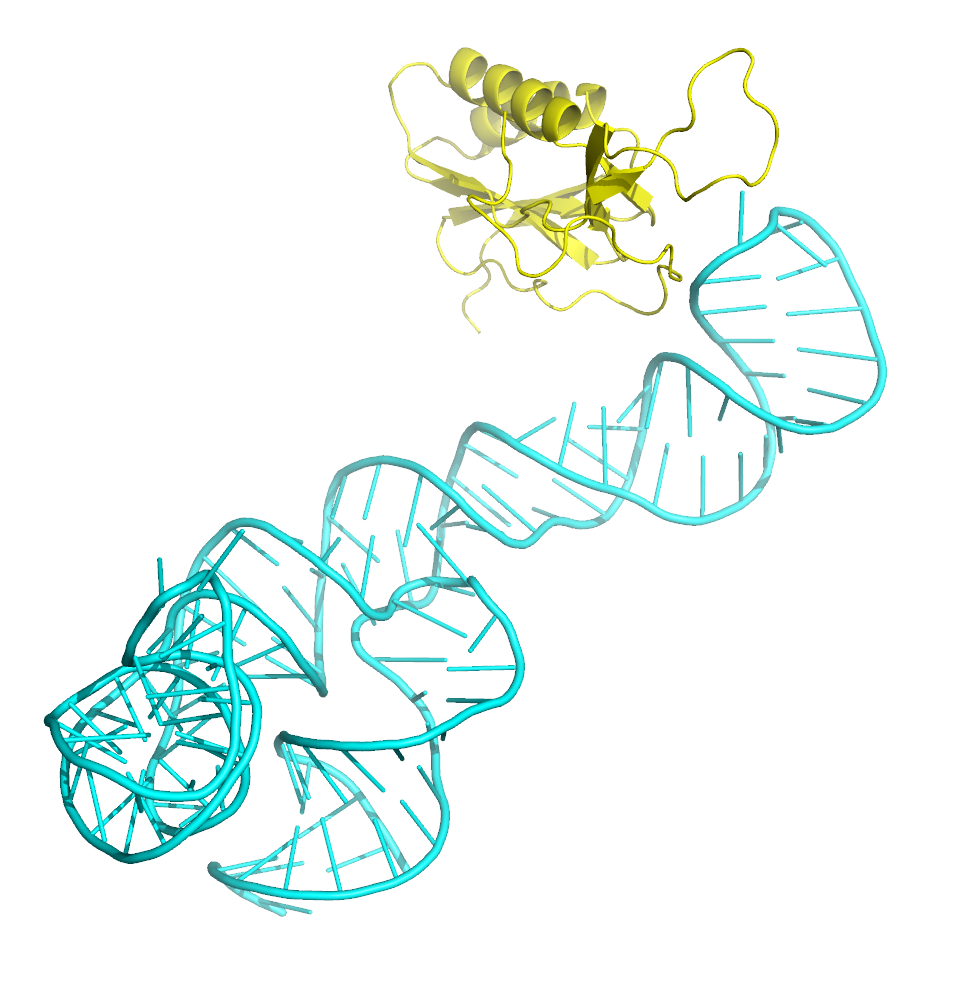}
    }\\
    
    \subfigure[Pose 3]{
       \centering
        \includegraphics[width=0.23\textwidth]{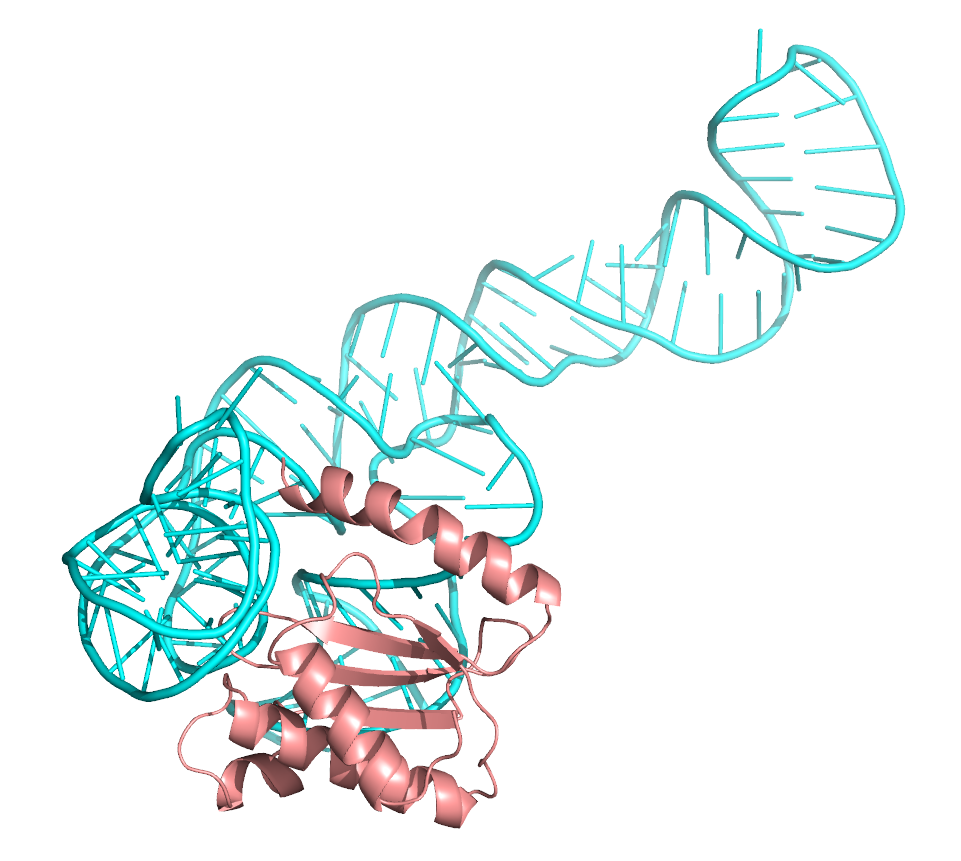}
    }\subfigure[Pose 4]{
        \centering
        \includegraphics[width=0.23\textwidth]{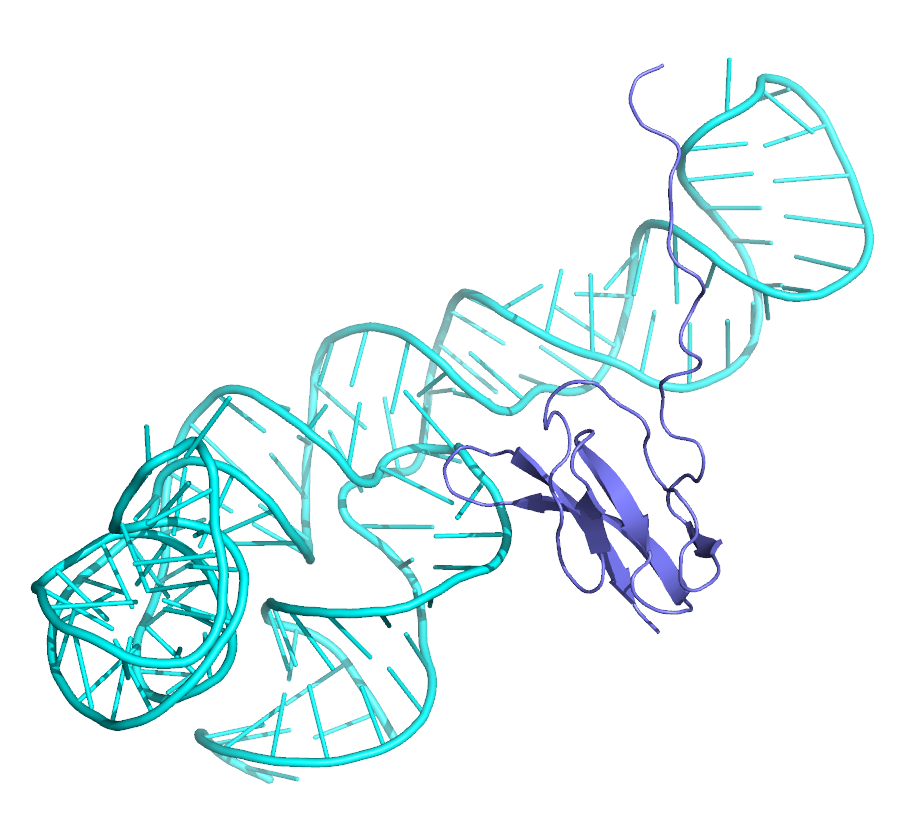}
    }
    
    \caption{An example of multi-pose interaction.}
    \label{fig6}
\end{figure}

\paragraph{Multi-pose interaction samples.}
In our RPI30k dataset, a protein can interact with an RNA with multiple poses, modeling these multiple poses as a node distance prediction task (as we've done in the MIDM task) can help \modelname{} understand the multiple binding sites in protein and RNA. Figure \ref{fig6} shows a multi-pose interaction example. Note that there are in total 7 poses in 5WFK, we show 4 of them here as an example.

\section{D. Baseline Selection}
In this section, we will describe the baselines we take in this work for the downstream tasks. Some of the protein-RNA binding affinity prediction approaches are inaccessible, and some only provide a web server with input restriction. One of the main purposes of our work is to build a standard evaluation dataset and an open-source approach for protein-RNA binding affinity prediction.

\subsection{Domain specific models.}

There are many models for protein-RNA binding affinity prediction. Unfortunately, most of them are inaccessible. For example, PNAB \cite{yang2019pnab} and PRA-Pred \cite{harini2024pred} provide a webserver but are unable to return results; The local version of PRdeltaGPred \cite{hong_updated_2023} and PRA-pred \cite{harini2024pred} needs registration of x3dna, which is currently disabled for registration. We exhaustively searched the existing recent protein-RNA binding affinity prediction tools and found two methods available via their web servers, which are Pred-PRBA \cite{deng_predprba_2019} and DeepNAP \cite{pandey_deepnap_2024}. We evaluate their performance on the PRA 201 subset due to the restriction of their server. We also compare our model's performance with the FoldX Suite 5.0 \cite{delgado2019foldx} designed especially for protein-RNA binding affinity and affinity change prediction. We also use the mCSM-NA test set and its web server for protein-RNA affinity change prediction.

\paragraph{Pred-PRBA \cite{deng_predprba_2019}.} This work extracts the protein-RNA binding interface's sequence and structure information as the input features for gradient-boosting regression trees. Then, they divide the protein-RNA complexes into 6 classes and predict the binding affinity for each class. We use their web server to get the $\Delta G$ results. And our samples may appear in their training set.

\paragraph{DeepNAP \cite{pandey_deepnap_2024}.} DeepNAP is a recent work that implements deep learning for protein-RNA binding affinity prediction. They extract sequence information for protein and RNA with different strategies, and separately input the sequence features into a 1D convolution. Then, they design some interaction modules to support both $\Delta G$ and $\Delta \Delta G$ prediction.  We use their web server to get the $\Delta G$ and $\Delta \Delta G$ results. And our samples may appear in their training set.

\paragraph{FoldX Suite 5.0 \cite{delgado2019foldx}.} FoldX 5.0 is designed to model protein interactions with RNA and small molecules. It is an energy-based method and predicts the free energy of unfolding of target protein-RNA sequences. We download the package and use their core suite for $\Delta G$ and $\Delta \Delta G$ prediction, and only compare the correlation coefficient in $\Delta G$ prediction.

\paragraph{mCSM-NA \cite{pires2017mcsm}.} mCSM-NA is designed especially for predicting the effects of mutations on protein-nucleic acid interactions. They take a graph-based signature for protein-RNA complex representation and refine a reliable dataset for model training and evaluation. We use their web server to get the $\Delta\Delta G$ results.

\subsection{Machine learning models.} 
Following the baselines selected in DeepNAP and Pred-PRBA, we also select several machine learning based methods for $\Delta G$ prediction. However, we do not use the manually extracted features. Instead, we use the embeddings from the PLM and the RLM as the models' input, named LM-enhanced machine learning baselines. In total, we implemented four methods, including Linear Regression, Random Forest, SVR, and MLP.

\subsection{Other related baselines}
To make a more comprehensive comparison and benchmark the 
performance for further development of protein-RNA affinity methods. As the advanced methods in predicting protein-ligand binding affinity usually use geometric graph based methods, we selected several representative baselines as chosen in \cite{li2021structure}, including GCN \cite{kipf2016semi} and GAT \cite{velivckovic2017graph} with geometric enhancement. We also report the results of EGNN \cite{satorras2021n} and GVP \cite{jing2020learning}, as they've been commonly used in protein and RNA-related tasks. Finally, we compare our method with IPA and LM enhanced IPA \cite{jumper2021highly}, which is a strong encoder in protein-protein affinity prediction and their mutation effect prediction, as described in \cite{luo_rotamer_2023}. Here we will describe their implementation details. All the training settings are the same, including the optimizer and the training scheduler. We take the Adam optimizer with an initial learning rate of 3e-5, and we set a plateau scheduler with a minimum learning rate of 1e-6. The node patch size at the interface is 256, the same as that in \modelname{}.

\paragraph{GCN \cite{kipf2016semi} and GAT \cite{velivckovic2017graph}.} GCN and GAT are used as protein-ligand binding affinity prediction methods in GraphDTA \cite{nguyen2021graphdta}. To make these methods structure-aware, we follow GraphDTA to add the distance information of the complex interface to the edge attributes. These methods serve as baselines for classic graph-based models.

\paragraph{EGNN \cite{satorras2021n}.} As a representative of geometric deep learning methods, EGNN is simple yet efficient. Since it is equivariant to 3D rotations and translations, there are many applications of EGNN for various 3D geometric protein-related tasks. We implement EGNN with a full-atom interface geometry. The input of EGNN is the node type and position of atoms, denoted as $(s, X)$.

\paragraph{GVP \cite{jing2020learning}.} Graph vector perceptron is another geometric deep learning method that was widely used in both protein \cite{han2024hemenet} and RNA \cite{joshi2024grnade} applications, making it a suitable baseline for predicting protein-RNA binding affinity predictions. We follow the common practice of using GVP for protein and RNA encoders, providing them with full-atom geometry at the binding interface. Following \cite{hsu2022learning} and \cite{joshi2024grnade}, there are four features in GVP, denoted as $(S_n, V_n, S_e, N_e)$, representing the scalar and vector feature for nods and edges, respectively.

\paragraph{IPA \cite{jumper2021highly} and LM + IPA.} Invariant point attention is a key module in AlphaFold-2 for structure understanding and prediction. Meanwhile, there are many subsequent works that use IPA as a structure encoder for predicting protein-protein binding affinity and mutation effects \cite{luo_rotamer_2023}. We follow the common use of IPA for dealing with protein input. For RNA, we choose 4 atoms as the input geometric information, described in the Methods section. There are in total 26 node types, including 20 normal amino acids, 1 unknown amino acid, 4 normal base types, and 1 unknown base type. Since IPA is the best-performed baseline, we further replace the 1D embedding of IPA with the output of ESM-2 and RiNALMo, resulting in a stronger baseline with better performance. The improvement in performance demonstrates the LMs are strong sequence information encoders for the binding affinity tasks.

\section{E. Detailed experiment results on PRA}
In this section, we report more detailed results on PRA tasks. First, we delete the worst predicted samples for each fold and observe the model's performance gain. Meanwhile, we report the performance of full-param training, LoRA training, and fix-LM training. Finally, we report the mean and standard deviation of the five-fold cross-validation on PRA310 and PRA 201. By evaluating the performance and the standard deviation, we can have a comprehensive knowledge of the model's overall performance and robustness.

\paragraph{Deleting the worst performed samples.} According to \cite{hong_updated_2023}, they delete the worst predicted samples and watch the performance gain. We also use this strategy to compare the performance of our baseline models on PRA201, since some models can only predict the affinity for samples from PRA201. We report the five-fold mean person correlation as the representative. The results can be found in Table \ref{table 6}, where $r$ is the ratio of validation samples removed in each fold.

\begin{table}[ht]
    \centering
    \small

    \begin{tabular}{lcccc}
        \toprule
        Method & $r=0$ & $r=3\%$ & $r=15\%$ & $r=25\%$   \\
        \midrule
        FoldX &0.212&0.174&0.289&0.236\\
        GCN  & 0.201 & 0.196 & 0.112 & 0.163  \\
        GAT &0.262&0.273&0.294&0.218\\
        EGNN &0.241&0.277&0.320&0.267\\
        GVP &0.240&0.248&0.214&0.236\\
        DeepNAP &0.345&0.468&0.598&0.662\\
        PredPRBA  & 0.370 & 0.416 & 0.556 & 0.673 \\
        IPA &0.532&0.583&0.658&0.737\\
        CoPRA &\textbf{0.569}&\textbf{0.661}&\textbf{0.743}&\textbf{0.788}\\
        \bottomrule
    \end{tabular}
    \caption{PCC of different deletion ratios.}
    \label{table 6}
\end{table}

As we can see, not all the models are getting a PCC improvement. The models with better initial performance with $r=0$ probably benefit more from the worst sample deletion experiment. Sometimes the PCC might decrease first and then increase, this is because we calculate the absolute Pearson coefficient of each fold, and some models might have negative initial PCC, such as GCN. Remarkably, the performance improvement of DeepNAP and PredPRBA is impressive. But we suppose that deleting the worst samples will result in deleting their non-seen samples in our dataset and keeping the ones that might appear in their training set. This phenomenon suggests that their models' generalization ability is limited because we've known some of our test samples appear in their training set. Meanwhile, deleting the worst predicted cases benefits IPA and \modelname{} monotonously, indicating a robust performance and a consistent prediction within different validation sets.

\begin{table}[ht]
    \centering
    \small

    \begin{tabular}{lcccc}
        \toprule
        Method & RMSE$\downarrow$ & MAE$\downarrow$ & PCC$\uparrow$ & SPC$\uparrow$   \\
        \midrule
        \modelname{}(all) &1.399 & \textbf{1.122} & 0.557 &0.549\\
        \modelname{}(lora) & 1.445  & 1.167  & 0.542 & 0.535 \\
        \modelname{}(fix-LM) &\textbf{1.391}&1.129&\textbf{0.580}&\textbf{0.589}\\
        \bottomrule
    \end{tabular}
    \caption{Performance of different LM settings.}
    \label{table 7}
\end{table}

\paragraph{More experiments of model fine-tuning.}
Our main purpose is to design an approach for bridging the cross-domain pre-trained models for protein-RNA binding affinity prediction. Therefore, we choose a lightweight strategy by fixing the pre-trained model. The strong performance of \modelname{} shows the effectiveness and efficiency of our approach. We also provide more experiment results here with the PLM and RLM unfixed in both pre-training and fine-tuning stages for a comprehensive comparison, as shown in Table \ref{table 7}. The pre-training and fine-tuning settings are the same as that for \modelname{}. The RMSE and MAE of all-parameter training are comparable to our fix-LM strategy while fixing LMs will result in a higher PCC and SPC. Meanwhile, LoRA training performs worse than both methods. We suppose it may be because we simply use the same settings of that in \modelname{} without further hyperparameter selection.

\paragraph{The detailed results on PRA310 and PRA201.} We report the five-fold mean performance and the standard deviation on the PRA310 and PRA201 datasets. As we can see, the pre-training of \modelname{} not only improves the performance but decreases the standard deviation, making our model more robust when dealing with different data splits. Moreover, adding LM embeddings will probably result in a more stable prediction, indicating the representation ability of LMs. The standard deviation of errors is larger than that of correlations because predicting the exact value is harder than predicting the trend.

\begin{sidewaystable*}[htbp]
\setlength{\belowcaptionskip}{0.3cm}

\centering
\small
\begin{tabular}{l|ccc|cccc|cccc}
\Xhline{1pt}
\multirow{2}{*}[-0.5ex]{Method} & \multirow{2}{*}[-0.5ex]{Struc} & \multirow{2}{*}[-0.5ex]{Seq} & \multirow{2}{*}[-0.5ex]{LM} &\multicolumn{4}{c|}{PRA310}  & \multicolumn{4}{c}{PRA201}                                           \\ 

\multicolumn{1}{c|}{} & \multicolumn{3}{c|}{}  & \multicolumn{1}{c}{RMSE$\downarrow$} & \multicolumn{1}{c}{MAE$\downarrow$} & \multicolumn{1}{c}{PCC$\uparrow$} & \multicolumn{1}{c|}{SCC$\uparrow$} & \multicolumn{1}{c}{RMSE$\downarrow$}                    & \multicolumn{1}{c}{MAE$\downarrow$} & \multicolumn{1}{c}{PCC$\uparrow$} & SCC$\uparrow$           \\ \midrule
LM+LR & \ding{55} & \ding{51} & \ding{51} & 1.801(0.131) & 1.472(0.130) & 0.365(0.111) & 0.348(0.096) & 1.750(0.183) & 1.383(0.135) & 0.370(0.118) & 0.362(0.362) \\

LM+RF &\ding{55} & \ding{51} & \ding{51} & 1.561(0.197) & 1.248(0.150) & 0.418(0.081) & 0.457(0.065) & 1.569(0.223) & 1.252(0.189) & 0.437(0.072) & 0.467(0.049) \\
LM+MLP &\ding{55} & \ding{51} & \ding{51}& 1.688(0.144) & 1.388(0.132) & 0.412(0.094) & 0.428(0.081) & 1.638(0.154) & 1.282(0.126) & 0.403(0.127) & 0.412(0.123) \\
LM+SVR &\ding{55} & \ding{51} & \ding{51}&  1.506(0.153) & 1.209(0.140) & 0.475(0.103) & 0.489(0.101) & 1.476(0.245) & 1.192(0.209) & 0.454(0.108) & 0.456(0.123) \\ 
LM+Transformer &\ding{55} & \ding{51} & \ding{51} & 1.481(0.215) & 1.192(0.180)  & 0.489(0.103) & 0.485(0.131) & 1.433(0.209) & 1.172(0.177) & 0.492(0.101) & 0.487(0.111) \\
DeepNAP* \cite{pandey_deepnap_2024} &\ding{55} & \ding{51} & \ding{55}& - & - & - & - & 1.964(0.161) & 1.600(0.178) & 0.345(0.079) & 0.349(0.123) \\ 
\cmidrule{1-12}

PredPRBA* \cite{deng_predprba_2019} &\ding{51} & \ding{55} & \ding{55} & - & - & - & - & 2.238(0.567) & 1.695(0.398) & 0.370(0.099) & 0.316(0.094) \\
FoldX$^\dag$ \cite{delgado2019foldx} &\ding{51} & \ding{55} & \ding{55} &   - &  -  & 0.212(0.075) & 0.283(0.134) & - & - &  0.212(0.056)  &  0.268(0.112)  \\
GCN \cite{kipf2016semi}  &\ding{51} & \ding{55} & \ding{55}  & 1.705(0.212)  & 1.378(0.172) & 0.145(0.103) & 0.144(0.120)  & 1.631(0.233) & 1.322(0.202) & 0.201(0.161)   & 0.203(0.173)     \\
GAT \cite{velivckovic2017graph}  &\ding{51} & \ding{55} & \ding{55}  & 1.644(0.202)  & 1.337(0.167) & 0.238(0.108) & 0.174(0.106)  & 1.542(0.199) & 1.235(0.191) & 0.262(0.054) & 0.221(0.077)    \\
EGNN \cite{satorras2021n} &\ding{51} & \ding{55} & \ding{55}  & 1.634(0.209)  & 1.340(0.185)   &  0.226(0.070) & 0.212(0.140) & 1.639(0.229) & 1.345(0.217) &    0.241(0.081)   &    0.217(0.116)   \\
GVP \cite{jing2020learning} &\ding{51} & \ding{55} & \ding{55}  & 1.678(0.221)  & 1.361(0.186)   &  0.262(0.288) & 0.283(0.138) & 1.702(0.256) & 1.372(0.229) &    0.240(0.082)   &    0.305(0.127)   \\
IPA \cite{jumper2021highly} &\ding{51} & \ding{55} & \ding{55}  & 1.462(0.236) & 1.208(0.190) & 0.495(0.119) & 0.496(0.158) & 1.464(0.214) & 1.191(0.188) & 0.510(0.103) & 0.514(0.092)\\

LM+IPA &\ding{51} & \ding{55} & \ding{51} & 1.454(0.211) & 1.198(0.170)  & 0.514(0.106) & 0.496(0.141) & \underline{1.405(0.214)} & \underline{1.159(0.169)} & 0.532(0.096) & 0.507(0.115) \\ \cmidrule{1-12}

\modelname{} (scratch) &\ding{51} & \ding{51} & \ding{51}& \underline{1.446(0.220)} & \underline{1.188(0.201)} & \underline{0.522(0.075)} & \underline{0.520(0.082)} & 1.428(0.201) & 1.172(0.184) & \underline{0.534(0.075)} & \underline{0.526(0.065)} \\
\modelname{} &\ding{51} & \ding{51} & \ding{51}& \textbf{1.391(0.142)} & \textbf{1.129(0.123)} & \textbf{0.580(0.033)} & \textbf{0.589(0.045)} & \textbf{1.339(0.199)} & \textbf{1.059(0.145)} & \textbf{0.569(0.067)} & \textbf{0.587(0.043)} \\

\Xhline{1pt}

\end{tabular}

\caption{The mean performance of 5-fold cross-validation on the PRA310 and PRA201 datasets. Sequence-based and structure-based models are listed in the tables. * The works only provide a web server with input requirements, so we only test them on the PRA201 subset. $^\dag$ The FoldX prediction is the complex energy change whose absolute value is much larger, thus we only compare the correlation coefficient here. LM is ESM-2 + RiNALMo. The standard deviation can be found in the Appendix E.}
\label{table 8}
\end{sidewaystable*}

\section{F. Ablation study on mask ratios}
\begin{figure}[htbp]
    \subfigure[Correlation]{
       \centering
        \includegraphics[width=0.23\textwidth]{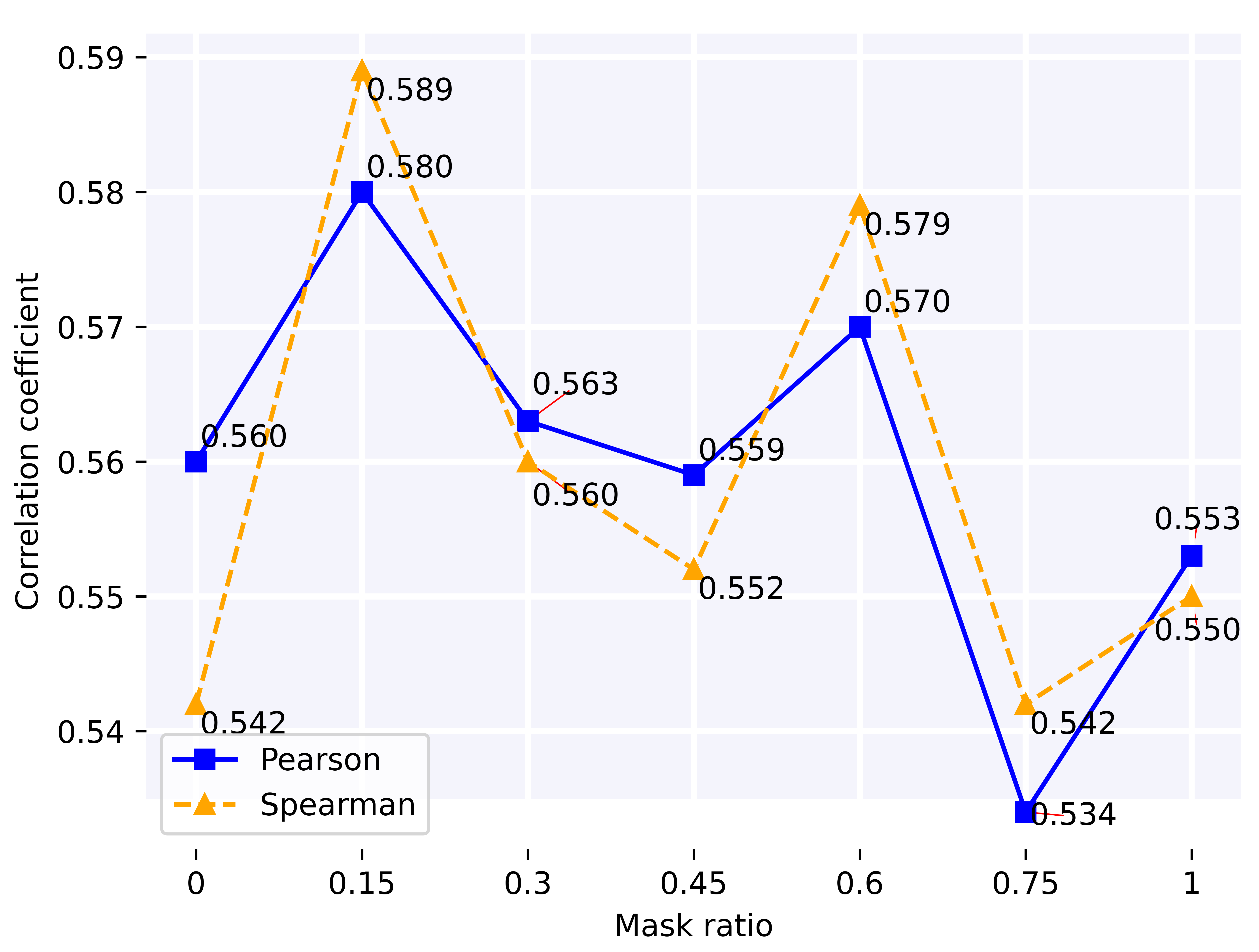}
    }\subfigure[Error]{
        \centering
        \includegraphics[width=0.23\textwidth]{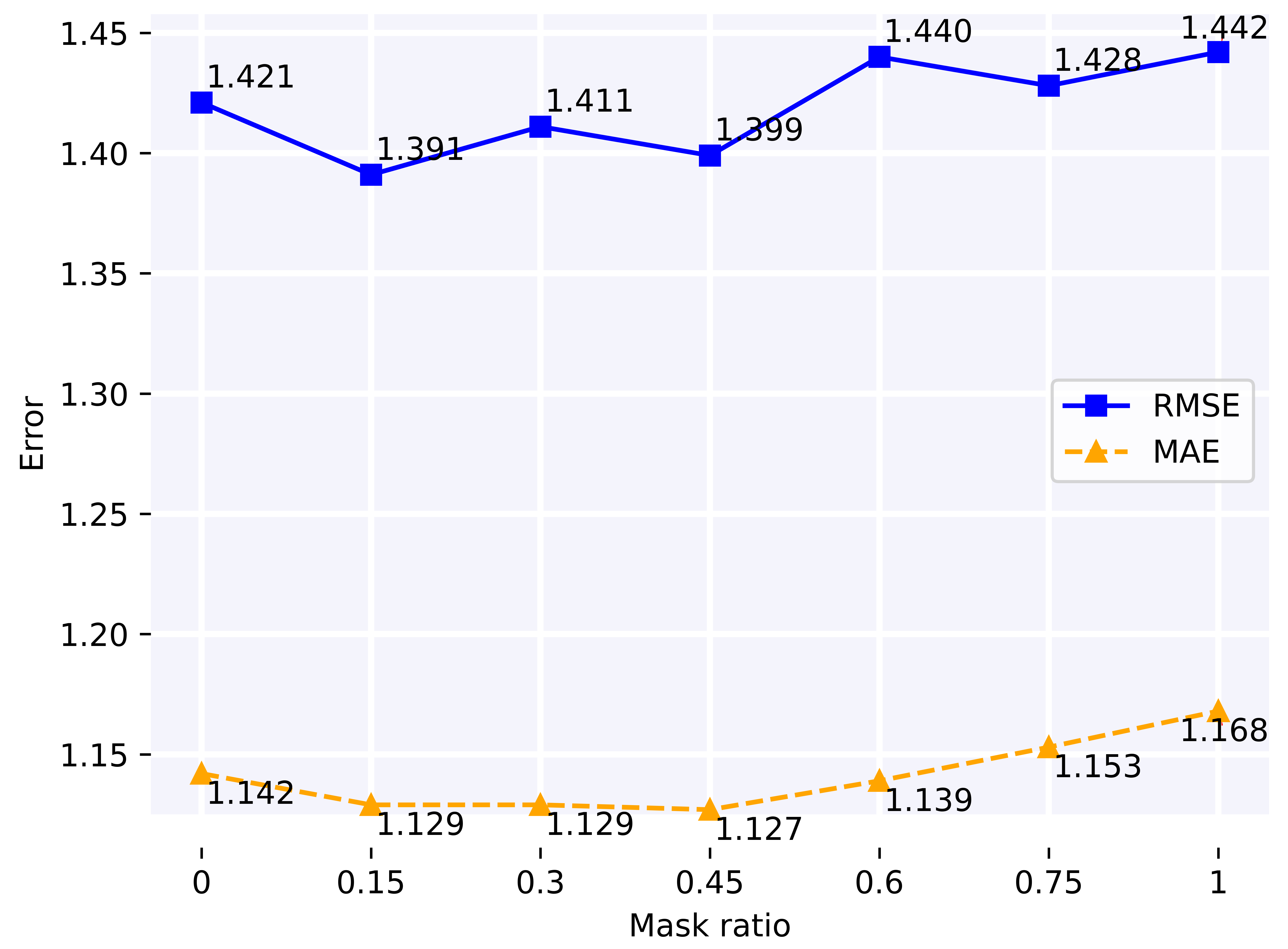}
    }\caption{Ablation study on mask ratio.}
    \label{fig7}
\end{figure}

In order to find the best mask ratio for the MIDM pre-train task, we experiment with different ratios, as shown in Figure \ref{fig7}, increasing the mask ratio will first increase the performance in all four metrics, because the improvement of task difficulty will help the model understand the interface distance better. As we can see, adding the mask interface modeling strategy will improve the performance initially, while when we increase the mask ratio, the overall performance will decrease in all the metrics. Because the higher mask ratio will cause a more corrupted pair embedding, making it too hard to train the model. Therefore, in \modelname{}, we set the final mask ratio as $15\%$, with the best overall performance.

\end{document}